\documentstyle[12pt,psfig,epsfig]{article}
\newcommand{\beq}{\begin{equation}}
\newcommand{\eeq}{\end{equation}}
\newcommand{\beqarray}{\begin{eqnarray}}
\newcommand{\eeqarray}{\end{eqnarray}}
\def\lsim{\raise0.3ex\hbox{$\;<$\kern-0.75em\raise-1.1ex\hbox{$\sim\;$}}}
\def\gsim{\raise0.3ex\hbox{$\;>$\kern-0.75em\raise-1.1ex\hbox{$\sim\;$}}}

\def\para{\vspace{0.3cm}\noindent} 

\def\cm{\,{\rm cm}}
\def\km{\,{\rm km}}
\def\m{\,{\rm m}}
\def\s{\,{\rm s}}  
\def\yr{\,{\rm yr}}

\def\mpc{\,{\rm Mpc}}

\def\erg{\,{\rm erg}} 
\def\ergs{\,{\rm ergs}} 
\def\ev{\,{\rm eV}}
\def\kev{\,{\rm keV}}
\def\mev{\,{\rm MeV}}
\def\gev{\,{\rm GeV}}
\def\tev{\,{\rm TeV}}

\def\sr{\,{\rm sr}}
\def\sec{\,{\rm sec}}

\def\lab{{\rm lab}}
\def\ob{{\rm ob}}

\def\lorentz{\Gamma_{300}}
\def\tvarob{{\Delta t^\ob}}
\def\tvarobsec{\left(\frac{\tvarob}{\rm sec}\right)}
\def\llow{L_{\rm L}}
\def\lhigh{L_{\rm H}}

\def\lhighlab{\lhigh^\lab}
\def\llowob{\llow^\ob}

\def\llowobn{L_{{\rm L},51}^\ob}
\def\rdlab{r_{\rm d}^\lab}
\def\Egamma{E_{\gamma}}
\def\Egammaob{E^{\ob}_{\gamma}}

\def\Egammamax{E_{\gamma,{\rm max}}} 
\def\Egammamaxob{E^{\ob}_{\gamma,{\rm max}}} 
\def\Egammamin{E_{\gamma,{\rm min}}}

\def\Egammath{E_{\gamma,{\rm th}}}
\begin{document}
\begin{center}
{\large \bf Detectability of Prompt TeV Gamma Rays of Proton 
Synchrotron Origin from Gamma Ray Bursts by AMANDA/ICECUBE Detectors\\}

\medskip

{Pijushpani Bhattacharjee\footnote{pijush@iiap.ernet.in}}\\
{\it Indian Institute of Astrophysics, Bangalore 560 034,
INDIA.}

\bigskip

{Nayantara Gupta \footnote{tpng@mahendra.iacs.res.in}}\\  
{\it Department of Theoretical Physics,\\
 Indian Association for the Cultivation of Science,\\
Jadavpur, Calcutta 700 032, INDIA.}  

\end{center}

\begin{abstract}
The detectability of a possible ``high'' energy ($>100\gev$) component 
of Gamma-Ray Bursts (GRBs) using AMANDA/ICECUBE large area muon 
detectors is examined within the context of a specific model of such high 
energy gamma ray 
production within GRBs, namely, the proton-synchrotron model, which 
requires protons to be accelerated to ultrahigh energies $\gsim10^{20}\ev$ 
within GRBs. In this model, the high energy component is distinct from, 
but may well be emitted in coincidence with, the usual ``low'' (keV--MeV) 
energy component observed by satellite-borne detectors. The AMANDA/ICECUBE 
detectors can detect TeV photons by detecting the secondary 
muons created by the TeV photons in the Earth's atmosphere. 
We calculate the muon signal to
noise ratio in these detectors due to TeV gamma-rays from individual 
GRBs for various assumptions on their luminosity, distance (redshift), 
Lorentz Gamma factor of the underlying fireball model, and various 
spectral characteristics  of the GRBs, including the effect of the 
absorption of TeV photons within the GRB as well as in the 
intergalactic infrared radiation background. The intergalactic absorption 
effect essentially precludes detection of TeV photons in the AMANDA 
detector for reasonable values of the luminosity in the high energy 
component, but they may well be detectable in the proposed ICECUBE 
detector 
which may have an effective area for downward-going muons a factor of 100 
larger than that in AMANDA. However, even in ICECUBE, only relatively 
close-by GRBs at redshifts $< 0.05$ or so can be expected to be detectable 
with any reasonable degree of confidence. We discuss the requirement on 
the luminosity of the GRB in the high energy component for 
its detectability in ICECUBE.      
\end{abstract} 
\newpage
\section{Introduction}
Gamma Ray Bursts (GRBs), short and intense bursts of photons 
observed mostly in the energy range of few tens of keV to few MeV  
with burst duration lasting from fraction of a second to several hundreds 
of seconds (see e.g., Ref.~\cite{fishman-meegan} for a review), are one of 
the most powerful and enigmatic astrophysical phenomena in the Universe. 
An afterglow phase, lasting from several hours to several days, has 
also been detected for many bursts. Detections of these afterglow phases 
in optical and/or x-rays and in some cases in radio, have provided 
valuable information regarding the nature of 
GRBs~\cite{grb_afterglow_rev}. For example, large 
redshifts of the host galaxies of some GRBs, measured 
from observations of their optical afterglows, have unambiguously 
established that GRB sources are extragalactic and indeed lie at 
cosmological distances. 

\para 
The total energy emitted in a GRB, estimated from 
the observed fluence and measured redshift (distance) of the burst source 
assuming isotropic emission, varies over a wide range from burst to burst, 
from about few $\times10^{50}\ergs$ to $\sim 10^{54}\ergs$. On the other 
hand, the material emitting the radiation could itself be highly 
collimated in the form of a jet, making the emitted radiation highly 
non-isotropic. 
One possible diagnostic, albeit not always a robust one, of 
a collimated emission is an expected break feature~\cite{rhoads} in the 
light curve of the optical afterglow of the GRB. Collimated emission 
significantly reduces the estimated total energy of the burst compared to 
that estimated assuming isotropic emission.  
From an analysis of the measured redshift and estimated 
collimation angle (from the break feature in the optical afterglow light 
curve) of some 18 bursts, Ref.~\cite{frail} finds  
a rather narrow distribution of the total energies of bursts with typical 
values around $5\times10^{50}\ergs$ with a FWHM of a factor of $\sim5$.  

\para 
The satellite-borne detectors, such as the BATSE on board the Compton
Gamma Ray Observatory (CGRO) and earlier detectors on board the Vela 
series of satellites, have detected GRBs mostly 
in the sub-MeV energy region. However, several GRBs with emission 
beyond 100 MeV, and in coincidence with the sub-MeV burst, have been 
detected~\cite{schneid,hurley,catelli} by the EGRET instrument on board 
the same CGRO, including the long-duration burst 
GRB940217~\cite{hurley} with emission extending to $\sim 18\gev$. 
This suggests the possibility that 
GRBs may, in addition to the sub-MeV photons, also emit much higher energy 
photons, perhaps even extending to
TeV energies as in some highly energetic Active Galactic Nuclei. For
power-law spectra falling with energy, the photon number flux at TeV
energies may be too low for these more energetic photons from GRBs to be 
detected by the satellite-borne detectors which have limited sizes. 
However, ground-based detectors can in
principle detect TeV photons from GRBs by detecting the secondary
particles comprising the ``air showers'' generated by
these photons in  the Earth's atmosphere. 

\para 
Indeed, three major ground-based gamma ray
detectors, the Tibet air shower array~\cite{tibet}, the HEGRA-AIROBICC
Cherenkov array~\cite{hegra} and the Milagro water-Cherenkov
detector~\cite{milagro} have independently claimed evidence, albeit not 
with strong statistical significance, for
possible TeV $\gamma$-ray emission from sources in directional and
temporal coincidence with some GRBs detected by BATSE. The estimated
energy in TeV photons in each case has been found to be about 1 to 2
orders of magnitude larger than the corresponding sub-MeV energies
measured by BATSE. More recently, a search~\cite{poirier} for sub-TeV 
($>10\gev$) gamma rays from GRBs with the Project GRAND array of muon 
detectors reported one possible detection at the 
$2.7\sigma$ level in coincidence with one BATSE GRB (971110), and an  
analysis~\cite{fragile} of the energetics of this event within the context 
of various possible models of TeV gamma ray emission from GRBs indicates   
a significantly higher estimated energy in TeV photons (by four orders of 
magnitude or more) than in sub-MeV photons.

\para 
Note that, since TeV photons are efficiently absorbed in the
intergalactic infrared (IR) background due to pair
production~\cite{stecker-dejager}, only relatively close by (i.e., low
redshift) GRBs, for which the absorption due to IR background is
insignificant, can be observed at TeV energies, which would explain the 
fact that only a few of the
BATSE-detected GRBs in the fields of view of the individual ground
detectors have been claimed to be detected at TeV energies.      

\para 
The significantly higher total energies estimated for the TeV 
photon emitting GRBs claimed to be detected by the ground-based 
detectors mentioned above raise the possibility that the TeV 
photons may constitute a separate `high' (GeV--TeV) energy component of 
GRB emission distinct from the `low' (keV--MeV) energy component detected 
by BATSE. For, otherwise, if the `observed' TeV photons came from the same 
low energy spectrum continued to TeV energies, then it would be difficult 
to understand the significantly higher total energy estimated for the 
bursts from the TeV observations than that estimated for the same bursts  
from sub-MeV observations by BATSE detector, unless the spectrum is hard 
with a differential power-law spectrum, $dN/dE\propto E^{-\alpha}$, with 
$\alpha<2$, which is not the case for the observed bursts. 

\para 
Confirmation of TeV gamma ray emission from GRBs 
would have major implications not only for the physics and astrophysics of 
GRBs as such, but also for several other phenomena. For example, 
some of the proposed mechanisms~\cite{vietri,bottcher-dermer,totani1,totani2}
of possible TeV photon emission from GRBs discussed 
below are directly linked with the proposed scenario of GRBs being 
possible sources of Ultrahigh Energy Cosmic Rays (UHECR)~\cite{grb_uhecr}. 
It has also been suggested~\cite{vazquez,totani3} that while the TeV
photons emitted by GRBs at large redshifts would be absorbed through
$e^+e^-$ pair production on the intergalactic infrared background, the
resulting electromagnetic cascades initiated by the produced pairs could
produce the observed extragalactic diffuse gamma ray background in the GeV
energy region. It is, therefore, important to study the detectability of 
the possible TeV photon emissions from GRBs in as many different ways as 
possible. 

\para
In this paper, we study the detectability of possible TeV photons from 
GRBs in the proposed ICECUBE~\cite{icecube} class 
underground muon detectors, within the context of a specific model of such 
high energy gamma ray production within GRBs, namely, the 
proton-synchrotron model discussed in detail below. The ICECUBE type 
muon detectors can detect TeV photons through detection of the secondary 
muons that are produced in Earth's atmosphere by high energy photons 
of energy above a few hundred GeV.  

\para 
The ICECUBE is the proposed full-scale 
version of the already operating AMANDA~\cite{amanda} detector. 
These are facilities primarily for detecting high energy 
neutrinos from cosmic sources. Located under the antarctic ice 
sheet, and consisting of arrays of photomultiplier tubes (PMTs) attached 
to strings and frozen in the deep ice, these detectors detect high energy 
neutrinos by detecting the muons produced by the neutrinos within the 
detector volume in the ice; the muons themselves are detected and their 
tracks (and thereby the arrival direction of the parent neutrinos) 
reconstructed using the array of PMTs which register the 
cherenkov light emitted by the muons traveling through the ice. In the 
final analysis, however, these are just muon detectors, and can, in 
principle, also detect the muons produced by TeV photons in the Earth's 
atmosphere.  
There is one basic difference, though, in the detection
method employed for photons as against that for neutrinos: While for
neutrinos one looks for upward-going muons in the detector, photons have
to be detected only through detection of the downward-going muons in the
detector.

\para 
The advantages of using these muon detectors over the conventional
air-Cherenkov telescopes for doing TeV gamma ray astronomy have been
expounded in Refs.~\cite{hsy,ah}. Compared to air-Cherenkov telescopes,
these detectors typically cover much larger fraction of the sky with 
large duty cycles. For example, the AMANDA detector  
covers more than a quarter of the sky with essentially 100\% efficiency. 
The detector is sensitive to muons with energies of a few hundred GeV 
and thereby to parent photons of energy of order a TeV and above. The 
fact that air-showers created by gamma rays are relatively ``muon poor'' 
(as compared to air-showers generated by hadrons) can be, to a large 
extent, compensated for by the relatively large effective area of a 
detector such as ICECUBE ($\sim 1\km^2$). In addition, 
as pointed out in Ref.~\cite{ah} (AH, hereafter), despite  
the otherwise large background of down-going cosmic-ray induced
atmospheric muons, the above muon detectors can in principle detect TeV 
gamma rays from transient sources like GRBs because, with the 
information on the time and duration of a burst provided by satellite
observation, the background is significantly reduced because it is 
integrated only over the relatively short duration of a typical GRB which 
is of order few seconds. 

\para 
Of course, possible TeV photons from GRBs can be detected by 
``conventional'' air-shower detectors, as exemplified by the claimed 
detections~\cite{tibet,hegra,milagro} already mentioned above. 
In this paper we focus on the AMANDA/ICECUBE detectors for
the advantages mentioned above and in particular for their potentially 
unique capability of simultaneously being detectors of neutrinos as well 
as photons of TeV energy and above, albeit not from the same source. 

\para 
In AH~\cite{ah}, the expected number of muons and the signal-to-noise 
ratios in AMANDA and Lake Baikal detectors due to TeV gamma-rays from 
individual GRBs were calculated for various different values of the GRB 
parameters. AH assumed a {\it single} power-law energy spectrum for 
individual GRBs continuing from the MeV BATSE 
band to TeV energies. However, as discussed above, it is 
more likely that the TeV photons constitute a separate 
high energy component distinct from the low energy ``BATSE'' component. In 
this paper, therefore, we consider such a model of TeV gamma ray 
emission from GRBs, namely, the proton-synchrotron model, in which, the 
TeV photons come from a new component different 
in origin than the sub-MeV ``BATSE'' photons. We study the detectability 
of the TeV photons in AMANDA/ICECUBE detectors within the 
context of this model, and study the sensitivity of the signal-to-noise 
ratio to various parameters such as the total luminosity of the GRB in the 
high energy (GeV--TeV) component, the redshift of the burst, etc., and 
in particular, the Lorentz Gamma factor ($\Gamma$) of the 
ultra-relativistic wind in the underlying fireball model (see below) of 
the GRB. 

\para 
The rest of this paper is organized as follows: In Sec.~2, we briefly 
discuss the motivation for considering the specific model of TeV photon 
production in GRBs in this paper, namely, the proton 
synchrotron model. The model is discussed in detail 
in Sec.~3. The internal optical depth of the produced high 
energy photons due to the process $\gamma\gamma\to e^+e^-$ within the 
GRB environment is calculated in detail in Sec.~4 giving the spectrum of 
high energy photons escaping from the GRB source. We also give the 
expression for the spectrum of photons arriving at Earth which includes  
the effect of the intergalactic absorption of TeV  
photons due to the process $\gamma\gamma\to e^+e^-$ taking place on the 
photons constituting the intergalactic infrared background.
In Sec.~5, the parametrization of 
the number of secondary muons in photon-induced showers in Earth's 
atmosphere used in this paper is discussed, and the muon 
signal-to-noise ratio in a ICECUBE class detector is calculated 
for individual GRBs as functions of various GRB parameters. The main  
results are discussed in Sec.~6, and conclusions are presented in Sec.~7. 

\section{Mechanisms of TeV Gamma Ray Production in GRBs: Motivation for 
Proton-Synchrotron Mechanism}
The fundamental source of the radiation from GRBs is thought to be 
the dissipation of the kinetic energy of ultra-relativistic (Lorentz 
factor $\Gamma\gsim {\rm few}\, 100$) bulk flow of matter caused by 
emission from a central engine, the nature of which is not yet known. 
Ultra-relativistic bulk flow of matter is obtained naturally in the 
fireball model (see. e.g., \cite{piran_revs} for reviews), in which the 
central engine initially releases a large 
amount of thermal energy $E_{\rm th}$ mostly in the form of an  
electron-positron-photon plasma in a small volume in a relatively baryon 
dilute medium. The baryon load --- consisting of the baryons emitted 
by the central engine as well as those present in the ambient medium --- 
is assumed to be small, i.e., 
$\eta^{-1}\equiv Mc^2/E_{\rm th}\ll 1$, where $M$ is the total mass of the 
baryon load. Because of the high 
optical depth of the photons due to the pair production process 
$\gamma\gamma\to e^+e^-$, the photons cannot escape; instead, the 
radiation pressure pushes the baryons into a shell of matter that expands 
outward. The initial thermal energy of the fireball thus gets transferred 
to the kinetic energy of an expanding shell of matter which, because of 
its small baryonic load, eventually attains Lorentz factor 
$\Gamma\sim\eta\gg1$ when essentially all of the initial thermal energy of 
the fireball is converted to kinetic energy of bulk flow. If the emission 
from the central engine is unsteady and sporadic, then the above process 
may result in multiple shells of baryons moving with different lorentz 
factors, whereas a single ultrarelativistic baryonic shell is obtained if 
the emission from the central engine is steady over the duration of the 
emission. In either case, the observed radiation is believed to 
arise from reconversion, through some dissipative process, of the kinetic 
energy of the ultra-relativistic bulk flow of matter back to internal 
energy at a large radius beyond the ``Thomson photosphere'' where the 
optical depth due to pair production, Thomson scattering and other 
processes is small.

\para  
The dissipation of kinetic energy of bulk flow may occur, in the case 
of unsteady emission, when different shells of matter moving with 
different $\Gamma$ collide with each other and give rise to ``internal 
shocks'', or, in the case of single 
impulsive emission, when the ultrarelativistic shell of matter 
decelerates as it sweeps up matter from the external medium, giving rise 
to an ``external shock''. The multiple baryon shells in the case of 
unsteady emission from the central engine would collide and eventually 
merge into one shell, which pushing into the external medium will also 
give rise to an external shock. Electrons as well as protons can be 
accelerated to high energies through Fermi mechanism at the internal and 
external shocks. In the multiple shell scenario, the prompt GRB phase is 
thought to result from the synchrotron radiation (and possibly also 
inverse-Compton scattering) of the electrons accelerated in the internal 
shocks while the afterglow phase results from the same process(es) 
occurring at the external shock. In the single shell scenario, on the 
other hand, there are no internal shocks, and both the prompt and 
afterglow phases result from processes at the external shock as it moves 
through the external medium. Although currently internal-external shock 
scenario is popular (see Ref.~\cite{piran_revs} and references therein), 
it has been argued~\cite{dermer-mitman} recently that the single impulsive 
external shock scenario can efficiently reproduce the observed 
properties of the prompt as well the afterglow phases of GRBs.  

\para
As mentioned in the previous section, a possible TeV component of GRB 
radiation is most likely to be of a different origin than the 
process that gives rise to the keV--MeV burst component. Mainly three 
processes have been suggested so far which can give rise to 
TeV photon emission from GRBs within the context of the fireball model of 
GRBs: (1) Synchrotron self-compton (SSC) process (see, 
e.g., \cite{dermer-chiang-mitman}), i.e, the inverse compton 
scattering of sufficiently high energy electrons on the ambient 
synchrotron photons within the source, the synchrotron photons being  
radiated by the same non-thermal electron population;  
(2) photo-pion production by ultrahigh energy protons 
within the source and the subsequent decay of the neutral pions to 
photons~\cite{waxman-bahcall97,rachen-meszaros98,muecke-etal99},  
and (3) synchrotron radiation by ultrahigh energy protons above 
$10^{20}\ev$ in the magnetic 
field within the source~\cite{vietri,bottcher-dermer,totani1,totani2}. 
In general, the required high energy non-thermal electrons and protons 
can be accelerated either in the internal shocks or in the external shock, 
and the associated TeV photons can be emitted either in the burst phase 
or the afterglow phase. In the present paper, we shall consider the  
case of prompt TeV photon emission in coincidence with the burst phase. 

\para
Both photo-pion production and the proton-synchrotron processes are 
motivated by the suggestion~\cite{grb_uhecr} that GRBs may be the sources 
of the observed ultrahigh energy cosmic rays (UHECR)~\cite{uhecr_revs} 
whose spectrum extends beyond $10^{20}\ev$. In this scenario, UHECR 
particles are assumed to be protons accelerated within GRBs. Thus, in this 
scenario, while the synchrotron radiation of electrons produces the 
observed low energy (keV--MeV) component, a high energy (GeV--TeV) 
component could be produced by the ultrahigh energy proton component 
either through synchrotron or photo-pion processes. 

\para
If one assumes that the total energy density of the high energy GeV--TeV 
component photons within the source is significantly higher than that of 
the low energy ``BATSE'' 
photons, as seems to be the case for the TeV photon emitting GRBs claimed 
to have been detected so far, and if one further assumes that the energy 
density of the magnetic field within the source is roughly in equipartition 
with the total energy density, then it can be shown~\cite{fragile}  
that, for production of photons of observed energy $\sim$ TeV and above, 
the proton-synchrotron process dominates over the photo-pion process, 
both processes being due to the same underlying non-thermal proton content 
within the source. On the other hand, if the SSC process were to produce 
the high energy component with a significantly higher energy density 
content than that of the low energy component, both components in this 
case being due to the same underlying non-thermal electron content within 
the source, then the energy 
density in the magnetic field would have to be significantly less than its 
equipartition value~\cite{fragile}, a condition that is at variance with 
models in which GRBs are sources of UHECR, which by and large require a 
magnetic field within GRB close to the equipartition 
limit~\cite{grb_uhecr}. 

\para
In this paper, we shall work within the framework of GRBs as sources of 
the observed UHECR up to $\sim10^{20}\ev$ and, based on arguments 
mentioned above, as far as the question of production of a high energy 
photon component of GRBs is concerned, we shall limit ourselves in the 
rest of this paper specifically to the proton-synchrotron model. 

\para 
The proton-synchrotron process may also play an 
important role in the internal energetics of GRBs. For example, 
it has been argued (see, e.g., ~\cite{totani1,totani2,totani3,totani4}) 
that, if the fundamental source of energy of the GRBs is indeed the 
kinetic energy of the ultrarelativistic bulk flow of matter as in the 
fireball model, then, at least initially, one would expect the total 
energy content in protons to be higher 
than that in electrons by a factor of $\sim m_p/m_e\sim2000$, where $m_p$ 
and $m_e$ are proton and electron rest mass, respectively. Immediately 
after dissipation of the kinetic energy of the bulk flow through 
formation of internal and external shocks, the total energy content of  
protons would, therefore, be higher than that of electrons. 
Transfer of energy from the protons to electrons through coulomb 
interactions is not efficient~\cite{totani4}. Instead, as suggested by  
Totani~\cite{totani4}, protons accelerated to energies above 
$10^{20}\ev$ may first produce TeV energy synchrotron photons which, in  
collision with the sub-MeV photons within the source, may produce $e^+e^-$ 
pairs through $\gamma\gamma\to e^+e^-$, thereby providing another channel 
through which energy can be transferred from the protons to electrons. If 
this process is very efficient in a GRB, then that GRB will not be a 
significant emitter of TeV photons. If on the other hand, the above pair 
production process is not very efficient, then the TeV photons would 
escape, resulting in strong TeV photon emission. The efficiency of the 
above pair production process for the TeV photons turns out to be a 
sensitive function of the initial bulk Lorentz factor $\Gamma$ (see 
below). Thus, small fluctuations in $\Gamma$ from burst to burst could 
give rise to wide variation of the TeV photon output from burst to burst 
even if the initial total energy input in all GRBs lies in a narrow 
range. Of course, the total energy output of a GRB in TeV photons will 
also depend on the shape of the accelerated protons' spectrum: If the 
proton spectrum is sufficiently hard so that most of the total energy in 
the proton component lies at the highest energy end of the spectrum where 
the synchrotron emission process is efficient, and if the pair-production 
process for the resulting TeV synchrotron photons is inefficient, then the 
bulk of the energy in the proton component (which is initially higher 
than that in the electron component) will escape from the source in 
the form of TeV photons, which may naturally explain the significantly 
higher total energy in the TeV photon component compared to that in the 
sub-MeV component estimated for the TeV photon emitting GRBs claimed to 
have been detected so far.   

\para 
Before going into quantitative discussions in the following sections, it 
is necessary to specify the relativistic reference frames used in the 
calculations below. There are three reference frames of 
interest: (1) The ``Wind Rest Frame'' (WRF), which is comoving 
with the average ultrarelativistic ``wind'' shell, (2) the ``lab" frame, 
in which the ultrarelativistic outflow moves with the average Lorentz 
factor $\Gamma$, and (3) the ``observer's frame" based on Earth. 
Quantities referred to in these frames will be indicated by the 
superscripts ``WRF", 
``lab" and ``ob", respectively. The ``ob" frame quantities differ from the 
corresponding ``lab" frame quantities only due to the cosmological 
redshift 
of the GRB source. Throughout this paper, physical quantities without 
any superscript for the reference frame will be understood to be 
referring to the WRF, unless explicitly stated otherwise. 

\section{The Proton-Synchrotron Model of TeV Photon Production in GRBs}
\subsection{The magnetic field strength in the wind rest frame}
We assume that there is a magnetic field ($B$) within the GRB environment, 
the strength of which is determined by the condition that the total energy 
density in the system is approximately equipartitioned between the 
magnetic field energy density, $B^2/8\pi$, and total particle-radiation 
energy density, $U$. 
Following Totani~\cite{totani1,totani2,totani3,totani4} and 
Ref.~\cite{fragile} we assume that $U\sim U_p\sim (m_p/m_e)U_\gamma$, 
where $U_\gamma$ is the energy density of the low energy BATSE photon 
component (assumed to arise mainly through electron synchrotron 
radiation), and $U_p$ is the energy density in the form of protons.      
Note that all the above quantities refer to the WRF. We thus have 
\beq
\frac{B^2}{8\pi}=\xi_B U_\gamma \frac{m_p}{m_e}\,,\label{B_equi}
\eeq
where $\xi_B$ is the equipartition factor assumed to be of order unity. 
For a GRB at a given redshift $z$, the WRF energy density $U_\gamma$
on the right hand side of Eq.~(\ref{B_equi}) can be expressed in terms of 
the observed variability time scale of the burst, $\tvarob$, and the 
observed luminosity, $\llowob$, in the low energy 
(keV--MeV) BATSE component inferred from the measured fluence and the 
known distance (redshift), in the following way: 
\beq
\llowob=\frac{\llow}{(1+z)^2}=\frac{1}{(1+z)^2}4\pi(\rdlab)^2
\Gamma^2cU_\gamma\,,\label{llowob_eq}
\eeq
where 
\beq 
\rdlab=2\Gamma^2c\tvarob/(1+z)\label{rdlab_eq}
\eeq
is the lab-frame value of the characteristic ``dissipation radius'' 
where the internal shocks 
are formed and from where most of the radiation is emitted. Note that 
$\rdlab$ is the value of the dissipation radius as inferred by the 
observer at Earth from the measured variability time scale after 
correcting for the time dilation due to expansion of the Universe.  

\para 
Note also that $\llowob$ as defined above is the effective $4\pi$ 
luminosity of the source assuming isotropic emission. If, instead, the 
source emits non-isotropically, for example, in a jet with opening solid 
angle $\Omega$, then the actual luminosity would be a factor 
$(\Omega/4\pi)$ smaller than $\llowob$. The same applies to the luminosity 
in the high energy component, $\lhigh$, used later. Obviously, our final 
results for the number of events per unit area in a detector on Earth and 
the associated signal-to-noise ratios are independent of whether or not 
the emission is isotropic, provided, of course, in the case of jetty 
emission, the emission jet from the source under consideration points 
toward Earth. Below we shall use the effective $4\pi$ luminosity in 
all our calculations.     

\para
Using equations (\ref{B_equi}) -- (\ref{rdlab_eq}), we can write 
\beq
B=6.8\times10^3\, \frac{\xi_B^{1/2}\left(\llowobn\right)^{1/2}(1+z)^2}
{\lorentz^3\tvarobsec}\,\, {\rm Gauss}\,,\label{B_value} 
\eeq
where $\lorentz\equiv\Gamma/300$, and 
$\llowobn\equiv\llowob/10^{51}\erg\s^{-1}$. 
\subsection{Synchrotron photon spectrum}
The synchrotron radiation photons emitted by protons of energy $E_p$ 
in a magnetic field of strength $B$ have the characteristic energy 
\beq
E_\gamma=\frac{3}{2}\frac{e\hbar 
BE_p^2}{m_p^3c^5}\langle\sin\alpha\rangle\,,\label{E_gamma1}
\eeq
where $e$ is the proton electric charge, and 
$\langle\sin\alpha\rangle=\pi/4$ is 
the average value of the sine of the pitch angle $\alpha$ of the proton.
Using Eq.~(\ref{B_value}) for the magnetic field, we get the relation 
between the proton energy and its synchrotron photon energy in the WRF: 
\beq
\frac{E_p}{\rm GeV}=\kappa_1\left(\frac{E_\gamma}{\rm 
GeV}\right)^{1/2}\,,\label{E_pE_gamma_rel_wrf}
\eeq
where 
\beq
\kappa_1=1.3\times 10^8\frac{\lorentz^{3/2}\tvarobsec^{1/2}}
{\xi_B^{1/4}\left(\llowobn\right)^{1/4}(1+z)}\,.\label{kappa1}
\eeq
Note that we can use the general relation 
\beq
E^\ob=\frac{\Gamma}{1+z}E\,\label{energy_trans}
\eeq
between energy in the observer's frame and and that in WRF, to express 
Eq.~(\ref{E_pE_gamma_rel_wrf}) in terms of observed energies: 
\beq
\frac{E_p^\ob}{\rm GeV}=\kappa_2\left(\frac{E_\gamma^\ob}{\rm 
GeV}\right)^{1/2}\,,\label{E_pE_gamma_rel_ob}
\eeq
where 
\beq
\kappa_2=2.3\times 10^9\frac{\lorentz^2\tvarobsec^{1/2}}
{\xi_B^{1/4}\left(\llowobn\right)^{1/4}(1+z)^{3/2}}\,.\label{kappa2}
\eeq
Note that $\kappa_2=\left(\frac{\Gamma}{1+z}\right)^{1/2}\kappa_1$. 

\para
Now, let 
\beq
\frac{dn_p}{dE_p}=A_p\left(\frac{E_p}{\rm GeV}\right)^{-\alpha_p}\, 
\label{p_spec1}
\eeq
be the differential spectrum of the protons 
(number density of protons per unit energy) accelerated 
in the internal shocks. The differential spectrum, $dn_\gamma/dE_\gamma$, 
of the synchrotron photons radiated by these protons is then related to 
the proton spectrum as 
\beq
f_{\rm ps}(E_p)E_p\frac{dn_p}{dE_p}dE_p = 
E_\gamma\frac{dn_\gamma}{dE_\gamma}dE_\gamma\,,\label{gamma_spec1}
\eeq
where $f_{\rm ps}(E_p)$ is the fraction of energy of a proton of 
energy $E_p$ lost to synchrotron radiation, and $E_p$ and $E_\gamma$ are 
related through Eq.~(\ref{E_pE_gamma_rel_wrf}). 

\para
The fraction $f_{\rm ps}(E_p)$ is just the fractional energy loss of 
the proton of energy $E_p$ through synchrotron radiation during one 
expansion timescale of the wind, i.e., 
\beq
f_{\rm ps}(E_p)=\frac{t_{\rm exp}}{t_{\rm 
ps}(E_p)}\,,\label{f_p_1} 
\eeq
where 
\beq
\frac{1}{t_{\rm ps}(E_p)} \equiv
\frac{1}{E_p}\left(\frac{dE}{dt}\right)_{\rm ps}(E_p)\,\label{t_ps_eq}
\eeq
is the inverse of the synchrotron loss time scale, with 
\beq 
\left(\frac{dE}{dt}\right)_{\rm ps}(E_p)\simeq 
c\, \frac{4}{9}\left(\frac{e^2}{m_pc^2}\right)^2
\left(\frac{E_p}{m_pc^2}\right)^2B^2\,\label{dEdt_sync}
\eeq
(assuming $E_p\gg m_pc^2$ and isotropic distribution of the pitch angles), 
and 
\beq
t_{\rm exp}\simeq\frac{1}{\Gamma}\frac{\rdlab}{c}\,\label{t_exp_eq}
\eeq
is the expansion time scale of the wind, both time scales being measured 
in the WRF.  

\para 
Since $t_{\rm ps}(E_p)\propto E_p^{-1}$, we see that protons of 
energy above a ``break'' energy $E_{pb}$ given by the condition 
\beq
t_{\rm ps}(E_p=E_{pb})=t_{\rm exp}\,,\label{E_pb_cond}
\eeq
will lose all their energy through synchrotron emission within one 
expansion time scale of the wind, thus giving 
\beq
f_{\rm ps}(E_p)=\left\{ 
\begin{array}{ll}\frac{E_p}{E_{pb}}\,, & \mbox{if 
$E_p < E_{pb}$}\,,\\
1\,, & \mbox{if $E_p \geq E_{pb}$\,.}
\end{array}
\right.\label{f_p_2}
\eeq
Using equations (\ref{t_ps_eq}), (\ref{dEdt_sync}) and (\ref{B_value}) 
we get  
\beq 
\frac{1}{t_{\rm ps}(E_p)} = 1.03\times10^{-11}
\frac{\xi_B\,\left(\llowobn\right)\,(1+z)^4}{\lorentz^6\tvarobsec^2}
\left(\frac{E_p}{\rm GeV}\right)\sec^{-1}\,,\label{t_ps_value}
\eeq
which, together with equations (\ref{E_pb_cond}), (\ref{t_exp_eq}) and 
(\ref{rdlab_eq}) gives  
\beq
E_{pb}\simeq 1.6\times10^8 \frac{\lorentz^5\tvarobsec}
{(1+z)^3\xi_B\left(\llowobn\right)}\,{\rm GeV}\,.\label{E_pb_value}
\eeq
Equations (\ref{gamma_spec1}), (\ref{p_spec1}), 
(\ref{E_pE_gamma_rel_wrf}), (\ref{kappa1}), (\ref{f_p_2}) and 
(\ref{E_pb_value}) then give the synchrotron photon spectrum in the WRF: 
\beq
\frac{dn_\gamma}{dE_\gamma}=\frac{1}{2}A_p\kappa_1^{-\alpha_p+2}
\left\{ 
\begin{array}{ll}\left(\frac{E_{\gamma b}}{\gev}\right)^{-1/2}
\left(\frac{E_{\gamma}}{\gev}\right)^{-\alpha_p/2-1/2}
 & \mbox{if $E_\gamma < E_{\gamma b}$}\,,\\
\left(\frac{E_{\gamma}}{\gev}\right)^{-\alpha_p/2-1}
 & \mbox{if $E_\gamma \geq E_{\gamma b}$\,,}
\end{array}
\right.\label{gamma_spec2}
\eeq 
where 
\beqarray
E_{\gamma b} & \equiv & \kappa_1^{-2}
\left(\frac{E_{pb}}{\gev}\right)^2\gev\,\\
 & = & 1.5
\frac{\lorentz^7\tvarobsec}
{\xi_B^{3/2}\left(\llowobn\right)^{3/2}(1+z)^4}\gev 
\,,\label{E_gamma_b_value}
\eeqarray
is the ``break" energy above which the synchrotron differential photon 
spectrum steepens by a power-law index 0.5 relative to the spectrum below 
that energy. 

\para
The spectrum (\ref{gamma_spec2}) cuts off above an energy 
$E_{\gamma,{\rm max}}$ given by 
\beq
E_{\gamma,{\rm max}}=\kappa_1^{-2}\left(\frac{E_{p,{\rm 
max}}}{\gev}\right)^2\gev\,,\label{E_gamma_max_eq}
\eeq
where $E_{p,{\rm max}}$ is the maximum energy of the protons accelerated 
in the internal shocks. The acceleration time scale is 
$t_{\rm acc}(E_p)\sim 2\pi\eta r_L(E_p)/c$, where $r_L(E_p)=E_p/(eB)$ 
is the Larmor radius and $\eta$ a factor of order unity.    
For reasonable ranges of values of the parameters  
of our interest, we have $t_{\rm acc} < t_{\rm exp} \leq t_{\rm ps}$ for 
$E_p\leq E_{pb}\,$, 
$t_{\rm acc} \leq t_{\rm ps} < t_{\rm exp}$ for $E_{pb} < E_p \leq 
E_{p,{\rm max}}\,$, and $t_{\rm ps} < t_{\rm acc}$ for $E_p > E_{p,{\rm 
max}}\,$. (Here the equality signs in the time scale conditions refer to 
the corresponding equality signs in the energy conditions.)  
Thus, $E_{p,{\rm max}}$ is in general larger than $E_{pb}$, and is 
determined by synchrotron energy loss, i.e., by the condition 
\beq
t_{\rm acc}(E_{p,{\rm max}})\simeq 
t_{\rm ps}(E_{p,{\rm max}})\,,\label{E_p_max_cond}
\eeq

\para 
Equation (\ref{E_p_max_cond}), together with equations (\ref{t_ps_eq}), 
(\ref{dEdt_sync}) and (\ref{B_value}), gives 
\beq
E_{p,{\rm max}}=9.8\times10^8 \eta^{-1/2} 
\frac{\lorentz^{3/2}\tvarobsec^{1/2}}
{\xi_B^{1/4}\left(\llowobn\right)^{1/4}(1+z)}\gev\,,
\label{E_p_max_value}
\eeq 
and consequently, 
\beq
E_{\gamma,{\rm max}}=54.73 \eta^{-1}\gev\,.\label{E_gamma_max_value} 
\eeq
In the observer's frame on Earth, equations (\ref{E_p_max_value}), 
(\ref{E_gamma_max_value}) and (\ref{energy_trans}) give 
\beq
E^\ob_{p,{\rm max}}=3\times10^{11}\eta^{-1/2}
\frac{\lorentz^{5/2}\tvarobsec^{1/2}}
{\xi_B^{1/4}\left(\llowobn\right)^{1/4}(1+z)^2}\gev\,,
\label{E_p_max_ob_value}
\eeq
and
\beq
E^\ob_{\gamma,{\rm max}}=1.64\times10^4 
\eta^{-1}\left(\frac{\lorentz}{1+z}\right)\gev\,.
\label{E_gamma_max_ob_value} 
\eeq
In the numerical calculations below, we shall take $\eta=1$. Note that 
equation (\ref{E_p_max_ob_value}) shows that, depending on values of the 
various parameters involved, protons may in principle be accelerated to 
(observed) energies above $10^{20}\ev$ in GRBs.  
\section{Optical Depth of High Energy Photons due to $e^+e^-$ pair 
Production and the Spectrum of Photons at Source and at Earth}
The spectrum, eq.~(\ref{gamma_spec2}), of the high energy photons produced 
by the proton synchrotron mechanism, has to be corrected for optical depth 
of the produced photons within the source (``internal optical depth'') in 
order to calculate the 
spectrum of the high energy photons emerging from the GRB source. The 
emergent spectrum so calculated has to be then further corrected for 
optical depth of the emitted high energy photons in the intergalactic 
medium (``external optical depth'') on their way from the source to Earth, 
in order to calculate the spectrum of photons arriving at Earth.  

\para 
The dominant source of the optical depth of high energy photons within the 
GRB environment as well as in the intergalactic medium 
is $e^+e^-$ pair production due to collision of the high energy photons 
with other photons in the medium.   
A test photon of energy $E_t$ can produce an $e^+e^-$ pair in a 
collision with a target photon of energy $\epsilon$ greater than a certain 
threshold value $\epsilon_{\rm th}$ given by 
\beq
\epsilon_{\rm th}=\frac{2(m_ec^2)^2}{E_t(1-\cos\theta)}\,,\label{pair_th}
\eeq
where $\theta$ is the angle between the two photons' directions of 
propagation. The cross section for this process for target photon energies 
$\epsilon$ above the threshold is~\cite{pair_cross_sec_ref}  
\beq
\sigma_{\gamma\gamma}(E_t,\epsilon,\theta) 
=\frac{3}{16}\sigma_T(1-\beta^2)\left[(3-\beta^4)
\ln\frac{1+\beta}{1-\beta}-2\beta(2-\beta^2)\right]\,,\label{pair_cross_sec}
\eeq
where $\sigma_T$ is the Thomson cross section and 
$\beta=[1-(\epsilon_{\rm th}/\epsilon)]^{1/2}$ is the center-of-mass 
speed of the outgoing pair particles in units of $c$. 

\subsection{Internal Optical Depth and the Lab Frame Spectrum of Photons 
Emerging from the Source} 
Let us first calculate the internal optical depth. 
The mean free path, $\ell_{\gamma\gamma}$, of the test photon of energy   
$E_t$ within the GRB source can be written, in the WRF, assuming isotropic 
distribution of the photons in the WRF, as~\cite{gould-schreder} 
\beq
\ell_{\gamma\gamma}^{-1}(E_t)=\frac{1}{2}\int d(\cos\theta)(1-\cos\theta)
\int_{\epsilon_{\rm th}(E_t,\cos\theta)}^\infty d\epsilon\,  
\frac{dn}{d\epsilon}\, \sigma_{\gamma\gamma}\,,\label{mfp_eq}
\eeq
where $dn/d\epsilon$ is the differential number density of the target 
photons (in the WRF) within the GRB source. 

\para
From Eq.~(\ref{pair_th}), we see that the lowest target photon threshold 
energy (which occurs for $\cos\theta=-1$) in the WRF, for a 
given test photon of observed energy $E_t^\ob=(\Gamma/(1+z))E_t$, 
is 
\beq
\epsilon_{\rm th, min}=250 \left(\frac{300\gev}{E_t^\ob}\right)
\frac{\lorentz}{1+z}\ev\,,\label{pair_th_min_value}
\eeq
which, in the observer's frame takes the form
\beq
\epsilon_{\rm th, min}^\ob=75 \left(\frac{300\gev}{E_t^\ob}\right)
\left(\frac{\lorentz}{1+z}\right)^2\kev\,,\label{pair_th_min_ob_value}
\eeq
From the above two equations, it is clear that, for the high energy 
photons of our interest, the target photons for pair production within the 
GRB source are mainly those that have observed energies typically in the 
low (keV--MeV) energy BATSE range. We shall, therefore, assume that 
the dominant contribution to the internal optical depth 
due to pair production for the high energy photons of our interest comes 
from target photons that constitute the observed low energy BATSE 
photon component. With this assumption, then, we can write 
\beq
\frac{dn}{d\epsilon}\, \propto \frac{d\phi^\ob}{d\epsilon^\ob}\,, 
\label{dn_depsilon_eq} 
\eeq
where 
$d\phi^\ob/d\epsilon^\ob$ is the differential spectrum of the observed 
BATSE photons, which, for most GRBs, is known to be well-fitted, within 
the BATSE energy range between $\epsilon_{\rm min}^\ob\approx 20\kev$ to 
$\epsilon_{\rm max}^\ob\approx 2\mev$, by 
a power-law with a break~\cite{band_spec}, 
\beq
\frac{d\phi^\ob}{d\epsilon^\ob}\, \propto \left\{
\begin{array}{ll}\left(\epsilon_b^\ob\right)^{\beta_l-\beta_h}
\left(\epsilon^\ob\right)^{-\beta_l}\,, & \mbox{if $\epsilon^\ob < 
\epsilon_b^\ob$}\,,\\
\left(\epsilon^\ob\right)^{-\beta_h}\,, & \mbox{if $\epsilon^\ob \geq 
\epsilon_b^\ob$}\,,
\end{array}
\right.
\label{batse_ob_spec}
\eeq 
where $\epsilon_b^\ob\approx 1\mev$ is the break energy of the observed 
spectrum, and $\beta_l\approx 1$ and $\beta_h\approx 2$. 

\para
The proportionality constant in (\ref{dn_depsilon_eq}) can be determined 
in terms of the observed luminosity $\llowob$ defined in equation 
(\ref{llowob_eq}) through the condition  
\beq
\int_{\epsilon_{\rm min}}^{\epsilon_{\rm max}}\epsilon\, 
\frac{dn}{d\epsilon}\, d\epsilon = U_\gamma = 
\frac{\left(1+z\right)^2\, \llowob}{4\pi 
c\left(\rdlab\right)^2\Gamma^2}\,, 
\label{batse_norm_cond}
\eeq
together with the relation (\ref{energy_trans}) between the energies in 
the WRF and observer's frame. 

\para
We can now perform the two integrals in equation (\ref{mfp_eq}) to 
calculate $\ell_{\gamma\gamma}^{-1}$. In doing this, it is 
convenient~\cite{gould-schreder} to change variables from $(\epsilon, 
\cos\theta)$ to $(\epsilon, s)$, with variable $s$ defined as 
\beq
s\equiv\frac{\epsilon E_t(1-\cos\theta)}{2(m_ec^2)^2}
=\frac{\epsilon}{\epsilon_{\rm th}}\equiv s_0z\,,
\label{s_def}
\eeq
with 
\beq
s_0\equiv\frac{\epsilon E_t}{(m_ec^2)^2}\,, \,\,\,\, 
z\equiv\frac{1}{2}(1-\cos\theta)\,.
\label{s_0_z_def}
\eeq
Note that $\beta=(1-1/s)$, so $\sigma_{\gamma\gamma}$ now becomes a 
function of the single variable $s$, allowing us to write equation 
(\ref{mfp_eq}) as 
\beq
\ell_{\gamma\gamma}^{-1}(E_t)=\frac{3}{8}\sigma_T
\left(\frac{m_e^2c^4}{E_t}\right)^2 
\int_{\frac{m_e^2c^4}{E_t}}^\infty d\epsilon\, \epsilon^{-2}\,   
\frac{dn}{d\epsilon}\, \varphi\left[s_0(\epsilon)\right]
\,,\label{mfp_eq2}
\eeq 
where 
\beq
\varphi\left[s_0(\epsilon)\right]=\int_1^{s_0(\epsilon)}\, s\, 
\tilde{\sigma}(s)\, ds\,, \,\,\, {\rm with} \,\,\, 
\tilde{\sigma}(s)\equiv\frac{16}{3}\frac{\sigma_{\gamma\gamma}}{\sigma_T}\,.
\label{varphi_def}
\eeq
The $\epsilon$ and $s$ integrals are now decoupled and are easily 
evaluated numerically. Finally, the internal optical depth $\tau_{\rm 
int}(E_t)$ for the high energy test photon of energy $E_t$ is given by the 
ratio of the wind expansion time scale $t_{\rm exp}=\rdlab/c\Gamma$ and 
the mean two-photon collision time scale 
$t_{\gamma\gamma}=\ell_{\gamma\gamma}/c$, giving 
\beq
\tau_{\rm int}(E_t)=\frac{\rdlab}{\Gamma}\ell_{\gamma\gamma}^{-1}\,.
\label{tau_int_eq}
\eeq
The dependence of $\tau_{\rm int}$ on $E_t$ and on various other 
parameters in the problem can be easily extracted if we make 
the simplifying approximation $\sigma_{\gamma\gamma}\simeq (3/16)\sigma_T
={\rm constant}$\footnote{Actually, for $s\gg 1$, $\sigma_{\gamma\gamma} 
\propto (\ln s)/s$, but the spectrum of target photons falls off rapidly 
for $\epsilon>\epsilon_{\rm th}$, so this approximation does not introduce 
large error.}. With this approximation, $\tilde{\sigma}(s)=1$, and  
$\varphi\left[s_0(\epsilon)\right]=(s_0^2-1)/2$, so the $\epsilon$ 
integral in equation (\ref{mfp_eq2}) can be performed analytically with 
the power-law form of $dn/d\epsilon$ given by equations 
(\ref{dn_depsilon_eq}), (\ref{batse_ob_spec}) and (\ref{batse_norm_cond}). 
For $\beta_l=1$, $\beta_h=2$, we get (in terms of quantities measured in 
the observer's frame) 
\beq
\tau_{\rm int}(E_t^\ob)\simeq \frac{3\sigma_T\llowob\mathcal{C}}{64\pi 
c^2\tvarob}
\left\{
\begin{array}{ll}
{\displaystyle  
\frac{(1+z)^4}{\Gamma^6}\frac{1}{3}\frac{E_t^\ob}{m_e^2c^4}\,,} & 
\mbox{for 
$E_t^\ob < E_{t,s}^\ob$}\,,\\
{\displaystyle 
\frac{(1+z)^2}{\Gamma^4}\left[\frac{1}{4\epsilon_b^\ob}+
\frac{1}{12\epsilon_b^\ob}\left(\frac{E_{t,s}^\ob}{E_t^\ob}\right)^2
+\frac{1}{2\epsilon_b^\ob}\ln\frac{E_t^\ob}{E_{t,s}^\ob}\right]\,,
}
& \mbox{for $E_t^\ob \geq E_{t,s}^\ob$}\,,
\end{array}
\right.
\label{tau_int_eq2}
\eeq
where 
\beq 
{\mathcal{C}}=\frac{1}{\displaystyle 
\left(\frac{\epsilon_b^\ob-\epsilon_{\rm 
min}^\ob}{\epsilon_b^\ob}\right)+\ln\left(\frac{\epsilon_{\rm 
max}^\ob}{\epsilon_b^\ob}\right)}\,,
\label{mathcal_C_def}
\eeq
and 
\beq
E_{t,s}^\ob=\left(\frac{\Gamma}{1+z}\right)^2
\frac{m_e^2c^4}{\epsilon_b^\ob} = 22.5 
\left(\frac{\lorentz}{1+z}\right)^2 
\left(\frac{1\mev}{\epsilon_b^\ob}\right)\gev\,,
\label{E_t_s_value}
\eeq
is the energy below which, as equation (\ref{tau_int_eq2}) shows, 
$\tau_{\rm int}\propto E_t^\ob$, while $\tau_{\rm int}(E_t)$ 
``saturates'' to a roughly constant value above this energy with  
two $E_t^\ob$ dependent 
correction terms one of which falls off as $(E_t^\ob)^{-2}$ while the 
other increases with $E_t^\ob$ only as $\ln E_t^\ob$. 

\para 
Equation (\ref{tau_int_eq2}) also shows the sensitive dependence of 
$\tau_{\rm int}$ on $\Gamma$. The results of our full 
numerical calculations of $\tau_{\rm int}$ shown in Figures 
1--4 clearly exhibit the expected dependence of $\tau_{\rm int}$ on 
various parameters in the problem. 

\para
With the internal optical depth calculated as above, the total number of 
high energy photons emitted by the GRB per unit time per unit energy as 
measured in the lab frame is given by 
\beq
{\displaystyle 
\frac{dN_\gamma^\lab}{dE_\gamma^\lab dt^\lab}=4\pi\left(\rdlab\right)^2c
\frac{dn_\gamma}{dE_\gamma}\exp\left(-\tau_{\rm int}(E_\gamma)\right)\,,
}
\label{dN_gamma_dt_dE_gamma_lab_eq}
\eeq
where $dn_\gamma/dE_\gamma$ is as given by equation (\ref{gamma_spec2}), 
and with $E_\gamma=E_\gamma^\lab/\Gamma$. The total photon luminosity in 
the high energy component emitted by the GRB source (due to proton 
synchrotron radiation), as measured by a lab frame observer, is   
\beq
\lhighlab\equiv\int E_\gamma^\lab \frac{dN_\gamma^\lab}{dE_\gamma^\lab 
dt^\lab}
dE_\gamma^\lab
= 4\pi\left(\rdlab\right)^2\Gamma^2c \int_{\Egammamin}^{\Egammamax} 
E_\gamma 
\frac{dn_\gamma}{dE_\gamma}\exp\left(-\tau_{\rm int}(E_\gamma)\right)
dE_\gamma\,,
\label{lhighlab_eq}
\eeq
where $\Egammamax$ is given by equation (\ref{E_gamma_max_value}).  
The exact value of $\Egammamin$ is not important as long as we take   
$\Egammamin < E_{\gamma b}$ since, the power-law index of the spectrum 
(\ref{gamma_spec2}) being less than 2 (for $\alpha_p<3$ which we always 
assume to be the case), the dominant contribution to the energy integral 
in 
the region $\Egammamin < E_\gamma \leq E_{\gamma b}$ comes from its value  
at $E_{\gamma b}$. 
We use equation (\ref{lhighlab_eq}) to eliminate the constant $A_p$ 
appearing in equations (\ref{p_spec1}) and 
(\ref{gamma_spec2}) in terms of $\lhighlab$ which we shall take as a free 
parameter. This completely fixes the lab frame emission spectrum of the 
high energy component given by equation 
(\ref{dN_gamma_dt_dE_gamma_lab_eq}). 

\subsection{External Optical Depth and the Spectrum of High Energy Photons 
at Earth}
Due to expansion of the Universe, the energies of individual 
photons are decreased by a factor of $(1+z)$ and time intervals are 
stretched by a factor of $(1+z)$, where $z$ is the redshift of the GRB 
source. Therefore, the spectrum of photons arriving at Earth (giving 
number of photons striking the top of the atmosphere per unit area per 
unit energy per unit time) is simply given by 
\beq
{\displaystyle 
\frac{dN_\gamma^\ob}{dA dE_\gamma^\ob dt^\ob}(E_\gamma^\ob)=\frac{1}{4\pi 
D_z^2}
\frac{dN_\gamma^\lab}{dE_\gamma^\lab dt^\lab}(E_\gamma^\lab)
\exp\left(-\tau_{\rm ext}(E_\gamma^\lab)\right)
}\,,
\label{dN_gamma_dA_dt_dE_gamma_ob_eq}
\eeq 
where, on the right hand side, $E_\gamma^\lab=(1+z)E_\gamma^\ob$, 
$\tau_{\rm ext}(E_\gamma^\lab)$ is the external optical depth, due 
to pair production through two-photon collision in the intergalactic 
medium, of a photon originating with energy $E_\gamma^\lab$ at the source, 
and $D_z$ is the radial coordinate distance of the GRB source at redshift 
$z$. 

\para   
For a spatially flat Universe (which we shall assume to be the case) with
$\Omega_\Lambda + \Omega_m = 1$, where 
$\Omega_\Lambda$ and $\Omega_m$ are, respectively, the
contribution of the cosmological constant and matter to the energy density
of the Universe in units of the critical energy density, $3H_0^2/8\pi G$, 
the radial coordinate distance $D_z$ is given by~\cite{mtw}  
\beq
D_z=\left(\frac{c}{H_0}\right)\int^{z}_{0}\frac{dz^\prime}
{\sqrt{\Omega_{\Lambda}+ \Omega_m(1+z^\prime)^3}}\,.\label{D_z_eq}
\eeq
Here $H_0$ is the Hubble constant in the
present epoch. In our calculations, we shall use the currently popular 
values  
$\Omega_\Lambda=0.7$, $\Omega_m=0.3$, and $H_{0}=65\, \km\, \sec^{-1}\,
\mpc^{-1}$.

\para
The main contribution to the external optical depth $\tau_{\rm ext}$ of 
GeV--TeV photons comes from $e^+e^-$ pair production due to collisions 
with the photons constituting the intergalactic infrared background, and 
has been studied by a number of authors. 
We shall use the optical depths given
in \cite{stecker-dejager}, where a parametrization has been given for
$\tau(\Egamma,z)$ for small $z$ appropriate for
our purpose. In our numerical calculations we use the optical depths given 
in Ref.~\cite{stecker-dejager} for the higher level of intergalactic 
infrared background in order to have conservative estimate of the number 
of muons produced in the detector. 

\section{TeV Photon Signal from GRBs in a ICECUBE Class Detector}
Photons of sufficiently high energy striking the earth's atmosphere 
interact with the air nuclei to produce 
pions; these photo-pions are the major source of muon production
by photons in atmospheric air-showers, among other possible physical
processes like direct pair production of muons by photons or
photo-produced charm decays. As first suggested by 
AH~\cite{ah}, it may be possible to detect the TeV photons from 
GRBs by detecting these 
(downward-going) muons created by the TeV photons in the Earth's 
atmosphere in AMANDA/ICECUBE detectors. Our aim in this paper is to 
explore this possibility within the context of a specific model of TeV 
gamma ray production in GRBs discussed in the previous sections.  

\subsection{Muons in photon-induced air-showers and muon signal-to-noise 
ratio}  
The number of muons 
with energy above $E_{\mu}$ in an air-shower produced by a single photon
of energy $\Egammaob$ striking the top of the atmosphere can be
parametrized~\cite{muon,ah}, for $E_\mu$ in the range 100 GeV to 1 TeV, by
the formula
\beq
N_{\mu}(\Egammaob,\,\,
\geq\!E_\mu)\simeq\frac{2.14\times10^{-5}}{\cos\theta}\frac{1}
{(E_\mu/\cos\theta)}\frac{\Egammaob}{(E_\mu/\cos\theta)}\,,
\label{muon_single_photon} 
\eeq
where $\theta$ is the zenith angle of the photon source, and all energies
are in TeV units. 
The above parametrization is valid for $\Egammaob/E_\mu\geq10$ \cite{ah}. 

\para 
The total number of muons with energy in excess of a threshold muon 
energy, $E_{\mu,{\rm th}}$, in a detector of effective area $A_{\rm eff}$ 
due to a 
single GRB of redshift $z$ and observed duration $T^\ob$ can then be 
written as 
\beq
N_\mu(\geq\!E_{\mu,{\rm th}})=A_{\rm eff} T^\ob 
\int_{\Egammath^{\ob}}^{\Egammamaxob}d\Egammaob 
\frac{dN_\gamma^\ob}{dA dE_\gamma^\ob dt^\ob} 
N_{\mu}(\Egammaob,\,\,\geq\!E_{\mu,{\rm th}})\,,\label{muon_all}
\eeq
where $\Egammath^{\ob}\simeq{10\times E_{\mu,{\rm th}}/\!\cos\theta}$ is 
the
minimum photon energy needed to produce muons of energy 
$E_{\mu,{\rm the}}$ in the
atmosphere~\cite{ah}, $\theta$ being the zenith angle of the photon
source. 

\para
Equation (\ref{muon_all}) constitutes our signal, 
$S=N_\mu(\geq\!E_{\mu,{\rm th}})$. This 
must then be compared to the background of atmospheric muons created by 
cosmic rays in the Earth's atmosphere. The differential spectrum of 
cosmic ray produced muons is given by~\cite{gaisser_book}
\beq
\frac{dN_{\mu,{\rm B}}}{dAdtd\Omega dE_\mu}\approx 
\frac{0.14E_{\mu}^{-2.7}}{\cm^2\s\sr\gev}
\left\{
{
\displaystyle
\frac{1}{1+\left(\frac{1.1E_\mu\cos\theta}{115\gev}\right)}+ 
\frac{0.054}{1+\left(\frac{1.1E_\mu\cos\theta}{850\gev}\right)}
}
\right\}\,.
\label{mu_bg_gaisser}
\eeq
The number of background muons in the detector is, therefore, 
\beq
B\equiv N_{\mu,{\rm B}}(\geq\!E_{\mu,{\rm th}})=A_{\rm eff}T^\ob 
\Delta\Omega 
\int_{E_{\mu,{\rm th}}}^\infty  
\frac{dN_{\mu,{\rm B}}}{dAdtd\Omega dE_\mu} dE_\mu\,,
\eeq
where we have included in the background the contribution only from the 
observed duration of the burst and from within a solid angle 
$\Delta\Omega=(\pi\delta\theta/180)^2$ around the direction of the burst, 
$\delta\theta$ being the angular resolution (in degrees) of the detector. 
For AMANDA/ICECUBE type detector, $\delta\theta$ is typically a few 
degrees.
We then define the signal-to-ratio, $S/N$, as 
\beq
\frac{S}{N}\equiv \frac{S-B}{\sqrt{B}}\,.
\eeq

\subsection{Muon threshold energy and effective area of detector for 
downward-going muons} 
For a muon to be detected by ICECUBE, it must have an energy 
$\gsim100\gev$\cite{icecube} within the detector volume. However, unlike 
in the case of the neutrino-induced muons which can be created by the 
neutrinos within or near the detector volume, the gamma ray 
induced muons in the atmosphere will have to have sufficient energy at 
the surface to reach the detector volume deep within ice (at a depth of 
$\gsim$ 1 km) and still have enough energy left to satisfy the trigger 
for the detector. This requires~\cite{referee}, for 
ICECUBE, a threshold muon energy of $E_{\mu,{\rm th}}\gsim250\gev$ at the 
surface for a vertical muon. In this paper we take $E_{\mu,{\rm 
th}}=250\gev$ in our numerical calculations of the muon 
signal-to-noise ratio for ICECUBE.  

\para 
We also need to specify what value to take for 
the effective area of the AMANDA/ICECUBE detectors for downward-going 
muons. As is well-known, these detectors are primarily designed to 
detect the upward-going muons created in the 
antarctic ice sheet by upward-going neutrinos from 
astrophysical sources, the main source of background for which is the 
upward-going muons due to upward-going atmospheric neutrinos.  
Most of the PMTs constituting the optical modules (OMs) are, therefore, 
placed downward-facing, while only a small number of PMTs are kept 
upward-facing. The AMANDA-B10 detector~\cite{amanda2}, for example, has a 
total of 302 OMs, in 295 of which the PMTs are downward-facing, while 7 
OMs have upward-facing PMTs. However, because of scattering of light in 
the ice surrounding the OMs, the OMs are not completely blind to light 
from its backward hemisphere. This effect makes each OM a 
relatively more isotropic light sensor in ice than in vacuum. 
For AMANDA-B10 detector, Ref.~\cite{amanda2} estimates the 
effective relative sensitivity of an OM to be $\sim$ 67\% in the forward 
hemisphere and $\sim$ 33\% in the backward hemisphere. 
From this, we can estimate that the effective area of the AMANDA-B10 
detector for downward-going muons would be $\sim 1/2$ of that for 
upward-going muons.          

\para
The above discussion only serves to illustrate that although underice muon 
detectors like AMANDA/ICECUBE are designed to optimally detect the 
upward-going muons, the effective area of the detector for downward-going 
muons is still a significant fraction of its effective area for 
upward-going muons. 

\para 
Of course, the effective area for muon detection depends  
on the energy of the muon and also on the zenith angle ($\theta$) of the 
muon's direction of propagation. Figure 23 of Ref.~\cite{amanda2} shows 
that for the AMANDA-B10 detector, the effective area for upward-going 
($\cos\theta\approx -1$) muons is 
$\approx 9000 {\rm m}^2$ for muons in the energy range 100 -- 1000 GeV. 
The effective area then decreases with increasing $\cos\theta$ reaching 
a small value at $\cos\theta=0$ (horizontal direction). However, 
following the discussion in the previous paragraph, we expect that the 
effective area will increase again as we go above the horizon, i.e., as 
$\cos\theta$ increases from 0 toward 1 (directly downward-going), with 
the value of effective area roughly $1/2$ of the corresponding value 
at the corresponding angle below the horizon. 

\para
As will be clear from our results discussed in the next section, the 
effective area for AMANDA-B10, and for that matter even the 
AMANDA-II detector, is not large 
enough to detect the TeV photon signal from GRBs with any reasonable 
degree of confidence within the context of the 
proton-synchrotron model discussed above, unless the GRB is 
extraordinarily bright in TeV photons. We shall, therefore, consider 
their detectability in the proposed ICECUBE detector which may reach an 
effective area of $\sim 10^6 {\rm m}^2$ for upward-going muons (at 
$\cos\theta=-1$). For the purpose of illustrating the kind of numbers for 
the signal-to-noise ratio we may expect in a ICECUBE class detector, 
we may assume that the ratio of the effective area for down-going muons 
to that for upward-going muons in ICECUBE will be roughly similar to 
that in the AMANDA-B10. We shall, therefore, take a value of  
$\sim 5\times 10^5 {\rm m}^2$ as a fiducial value of the effective area 
for directly downward-going muons in ICECUBE for the muon energy range of 
our interest. Our results for the signal-to-noise ratio discussed in 
the next section can be easily 
scaled (as $\sqrt{A_{\rm eff}}$) for the actual effective area.  

\section{Results and Discussions}
We now discuss our main results. There are 11 free parameters in 
the model. These are: $\llowobn,\, \xi_B,\, \beta_l,\, \beta_h,\, 
\epsilon^\ob_b,\, \tvarob,\, T^\ob,\, \Gamma,\, z,\, \lhighlab,\, 
{\rm and} \,  \alpha_p$. Out of these we fix 
$\llowobn=1\,$, $\xi_B=0.5\,$, $\beta_l=1\,$, $\tvarob=0.5\sec\,$, 
and $T^\ob=(1+z)10\sec\,$, and study the dependence of our various results 
on 
the other remaining free parameters. Scalings of our results with respect 
to the fixed parameters are obvious from the relevant equations given in 
the earlier sections.   

\para 
Figures 1--4 show how the 
internal optical depth due to pair production within the GRB source as a 
function of the observed energy depends on various parameters that 
characterize the burst. The functional dependence on energy is as expected 
from the approximate analytic expression for $\tau_{\rm in}$ given by 
equation (\ref{tau_int_eq2}), i.e., $\tau_{\rm in}$ increases with energy 
up to a saturation energy given by equation (\ref{E_t_s_value}) beyond 
which $\tau_{\rm in}$ is approximately constant. 

\para 
The most sensitive dependence 
of $\tau_{\rm in}$ is on $\Gamma$ as indicated by equation 
(\ref{tau_int_eq2}) and as shown in Fig.~1: Larger values of $\Gamma$ give 
smaller values of $\tau_{\rm in}$, and thus depending on a moderately 
large value of $\Gamma$ in the range, say, 300--500, we can have 
$\tau_{\rm in}\lsim 1$. 

\para
Fig.~2 shows how $\tau_{\rm in} (\Egammaob)$ depends on redshift of the 
burst. For a given value of the $\llowob$, higher redshift implies higher 
luminosity at the source. This in turn implies higher density of 
target photons for pair production inside the source, giving higher 
internal optical depth. Also, clearly, $\tau_{\rm in}$ depends on the 
spectrum of the target photons within the source. 
We show the dependence 
of $\tau_{\rm in} (\Egammaob)$ on two of the parameters that determine the 
target photon spectrum, namely, the break energy $\epsilon_b^\ob$ (Fig.~3) 
and the power-law index $\beta_h$ of the spectrum beyond the break 
(Fig.~4). 

\para
In Fig.~5 we display for illustration a typical predicted spectrum of the 
high energy component of a GRB due to proton-synchrotron process, 
where we also show the observed low energy ``BATSE'' spectrum for 
comparison. The effects of the internal and intergalactic optical depths 
due to pair production on the spectrum of the high energy component are 
also shown.  

\para
In Figures 6--8 we show how the signal-to-noise ratio in the detector 
depends on various parameters that characterize the burst. In all these 
figures, as also in Fig.~9, we assume an effective area of the detector 
for downward-going muons to be $\sim 5\times10^5\m^2$, assume the 
source to be at zero zenith angle, and take the threshold muon energy (at 
the surface) for detection by ICECUBE to be 250 GeV.
Figure 6 shows the dependence of the 
signal-to-noise ratio on the redshift of the GRB for three different 
values of the emitted luminosity in the high energy component at the 
source, $\lhighlab$. Note that $\lhighlab$, as defined in equation 
(\ref{lhighlab_eq}), is the effective $4\pi$ luminosity including the 
correction due to internal optical depth. It is clear that for reasonable 
values of other parameters, detectability of the high energy component for 
GRBs at redshifts beyond about 0.05 requires excessively high values of 
$\lhighlab$ upward of $\sim 10^{58}\ergs/\sec$. 
Dependence of the signal-to-noise ratio on the Lorentz factor $\Gamma$ of 
the underlying fireball model of the GRB is shown in Fig.~7. From Fig.~7, 
we see that TeV photons from a GRB at redshift $z\sim0.1$ is detectable 
with a signal-to-noise ratio of $\sim 10$ if 
$\lhighlab\sim5\times10^{57}\ergs/\sec$ and $\Gamma\simeq400$. 

\para
In the proton-synchrotron model under consideration, the high energy 
photon spectrum depends on the spectrum of the protons accelerated within 
the source. In Fig.~8, we show the dependence of the signal-to-noise ratio 
on the power-law index ($\alpha_p$) of the differential spectrum of the 
protons, for various values of the redshift of 
the GRB. As expected, a steeper proton spectrum (i.e., larger values of 
$\alpha_p$) gives a steeper synchrotron photon spectrum, which gives 
smaller number of high energy photons emitted from the source and hence 
smaller signal-to-ratio in the detector. This dependence on $\alpha_p$ is 
summarized in Fig.~9 where we show the behavior of the 
minimum luminosity in the high energy component, $\lhighlab$, 
required for detection with a signal to noise ratio of 5 or larger, as a 
function of $\alpha_p$, for various values of the 
redshift $z$ of the GRB. Clearly, as expected, a harder proton spectrum 
(i.e., smaller values of $\alpha_p$) gives better 
prospect for detectability of the TeV photons in the proton-synchrotron 
model. 

\para
From the above discussions while it is clear that, within the context of 
the proton synchrotron model, the energetic 
requirement for detectability of a possible high energy ($\gsim 
\tev$) photon component of GRBs in ICECUBE class detector is 
admittedly rather high, the required total energy estimates are not  
implausible, especially for reasonably close-by GRBs at redshifts $<0.05$ 
or so, as seen from Fig.~9. 
  
\para
We may compare the luminosity requirements discussed above for TeV 
photons with that for ultrahigh energy protons in the scenario where GRBs 
are sources of UHECR~\cite{grb_uhecr}. In order to explain the observed 
flux of UHECR, one needs a typical GRB to emit a total energy of $\sim 
{\rm few\, } \times 10^{53}\erg$ in UHE protons with energy $>10^{19}\ev$,  
assuming that GRB rate evolves with redshift like the star 
formation rate in the Universe, with a local ($z=0$) GRB rate of $\sim 
5\times10^{-10}\mpc^{-3}\yr^{-1}$\cite{bahcall-waxman_uhecr}. Note that 
the above number for the total energy in UHE protons refers to the 
total energy {\it escaping} from the GRB source in the form of UHE 
protons. The total energy in the UHE proton component produced {\it 
within} the GRB may be significantly higher, depending on the various 
energy loss processes of the UHE protons within the GRB source. If 
synchrotron radiation is the dominant energy loss process for the UHE 
protons within the GRB source, and if this process is very efficient, and 
further if the pair-production optical depth of the resulting GeV--TeV 
photons within the source is sufficiently small, then the escaping 
GeV--TeV photon luminosity of the source may even be larger than the 
escaping UHE proton luminosity. 

\para 
An indicator of the efficiency of the proton-synchrotron process is the 
value of the energy $E_{pb}$, given by equation (\ref{E_pb_value}), 
above which all protons lose all their energy through synchrotron 
radiation. Thus, smaller the value of this energy, higher is the 
synchrotron photon yield for a given spectrum of the protons. 
Specifically, if we assume that the internal pair-production optical depth 
of the high energy photons is negligible, then for a proton spectrum with 
index $\alpha_p=2$, it is easy to show that the condition that the 
escaping high energy photon luminosity (due to synchrotron radiation of 
protons) be higher than the escaping proton luminosity is 
\beq
E_{pb}< 2.72 \left(E_{p,{\rm min}} E_{p,{\rm max}}\right)^{1/2}\,,
\label{escaping_lum_cond}
\eeq
where $E_{p,{\rm min}}$ and $E_{p,{\rm max}}$ are the minimum and maximum 
energies of the protons, respectively. For appropriate choice of various 
parameters characterizing a GRB such as $\xi_B$ (measuring the internal 
magnetic field), $\Gamma$ and so on, the above condition may well be 
satisfied. 

\para 
Lastly, we remind the reader that throughout the above discussion, the 
luminosity values we discussed refer to the isotropic effective $4\pi$ 
luminosities; as already mentioned, for collimated emission the required 
luminosity values would be appropriately smaller by a factor of 
$\Omega/4\pi$, where $\Omega$ is the solid angle of the collimated jet. 

\section{Summary and Conclusions}
It is possible that GRBs emit not only sub-MeV photons as detected in
satellite-borne detectors, but also higher energy photons extending to TeV
energies. For a GRB photon spectrum falling with energy, as is usually the
case, the non-detection of TeV photons in satellite-borne detectors could 
be due to their limited size. However, as first 
suggested in Refs.~\cite{hsy,ah}, TeV photons from GRBs 
might be detectable in the existing under-ice muon detectors such as 
AMANDA and future detectors such as the proposed ICECUBE, by
detecting the muons in the atmospheric showers created by TeV photons.
In this paper, we have made detailed
calculations of the detectability of possible TeV photons from
individual GRBs by AMANDA/ICECUBE detectors
within the context of a specific model of such high energy gamma ray 
production within GRBs, namely, the proton-synchrotron model, which 
requires protons to be accelerated to ultrahigh energies $\gsim10^{20}\ev$ 
within GRBs. In this model, the high energy component is distinct from, 
but may well be emitted in coincidence with, the usual ``low'' (keV--MeV) 
energy component observed by satellite-borne detectors. 
We have calculated the expected number of muons and the signal to
noise ratio in these detectors due to TeV gamma-rays from individual 
GRBs for various assumptions on their luminosity, distance (redshift), 
Lorentz Gamma factor of the underlying fireball model, and various 
spectral characteristics  of the GRBs, including the effect of the 
absorption of TeV photons within the GRB as well as in the 
intergalactic infrared radiation background. The intergalactic absorption 
of TeV photons essentially precludes detection of TeV photons in the 
currently operating AMANDA detector for any reasonable values of the 
luminosity in the high energy component, but they may well be detectable 
in the proposed ICECUBE detector 
which may have an effective area for downward-going muons a factor of 100 
larger than that in AMANDA. However, even in ICECUBE, only relatively 
close-by GRBs at redshifts $< 0.05$ or so can be expected to be detectable 
with any reasonable degree of confidence. The required isotropic 
luminosity of the high energy photon component is upward of 
$10^{56}\ergs/\sec$ and are generally found to be more than 3--4 orders 
of magnitude higher than typical estimated isotropic luminosities in the 
keV--MeV BATSE energy band. Such high luminosities may not, however, be 
impossible in some GRBs depending on various parameters that characterize 
the GRB. 

\section{Acknowledgment}
One of us (NG) wishes to thank IIA, Bangalore for hospitality where a
major part of this work was done. PB acknowledges partial support under
the NSF US-India cooperative research grant \# INT-9714627.

\newpage 
\begin{figure}
\epsfig{figure=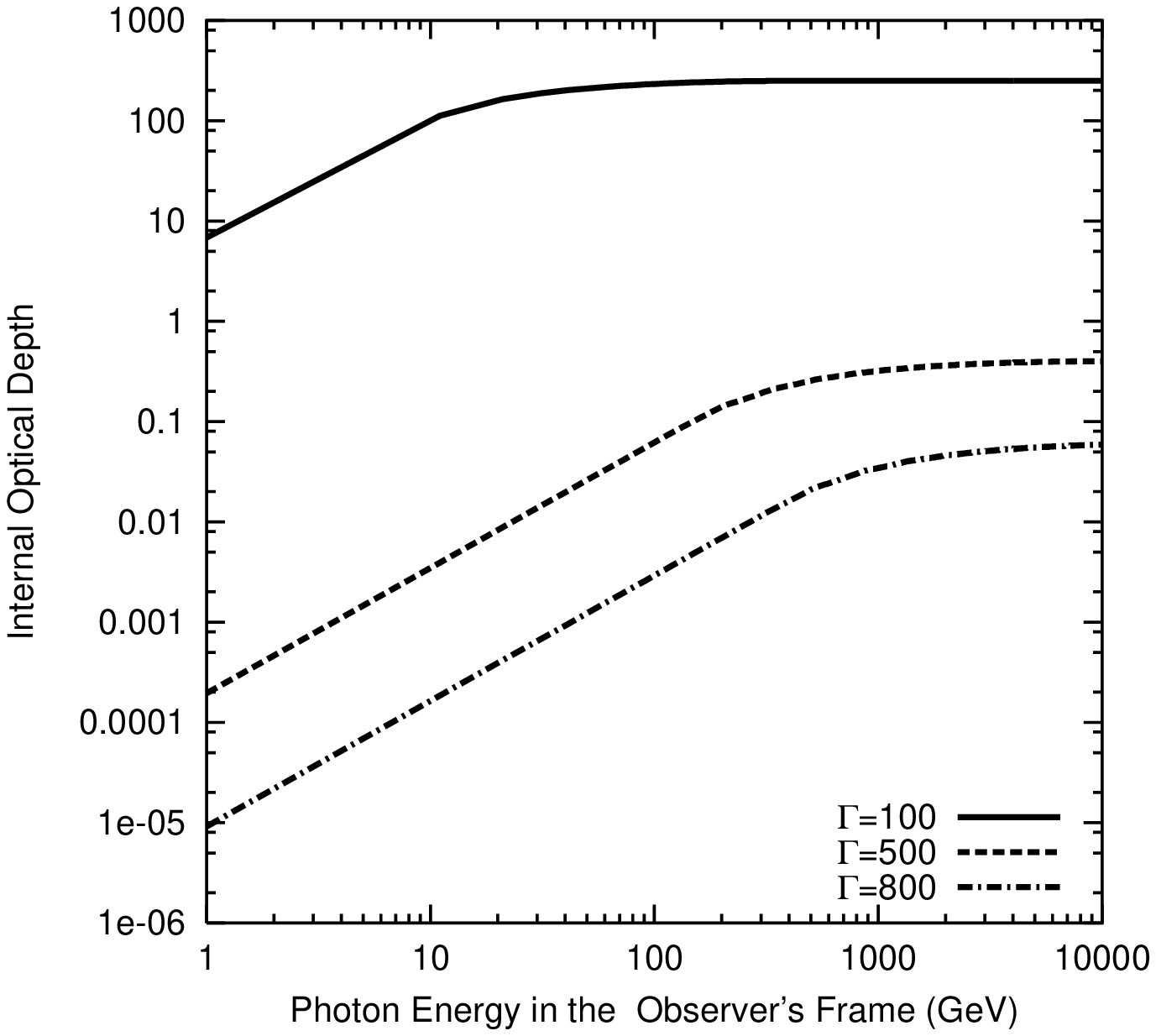,height=9.5cm} 
\caption{Internal optical depth as a function of the photon's energy in 
the observer's frame,  
for various values of $\Gamma$ as indicated. Values taken for other 
relevant parameters are: 
$\llowobn=1,\, \beta_l=1,\, \beta_h=2.25,\, 
\epsilon^\ob_b=0.5\mev,\, \tvarob=0.5\sec,\, {\rm and}\,\, z=0.1\,.$ 
}
\end{figure}

\newpage 
\begin{figure}
\epsfig{figure=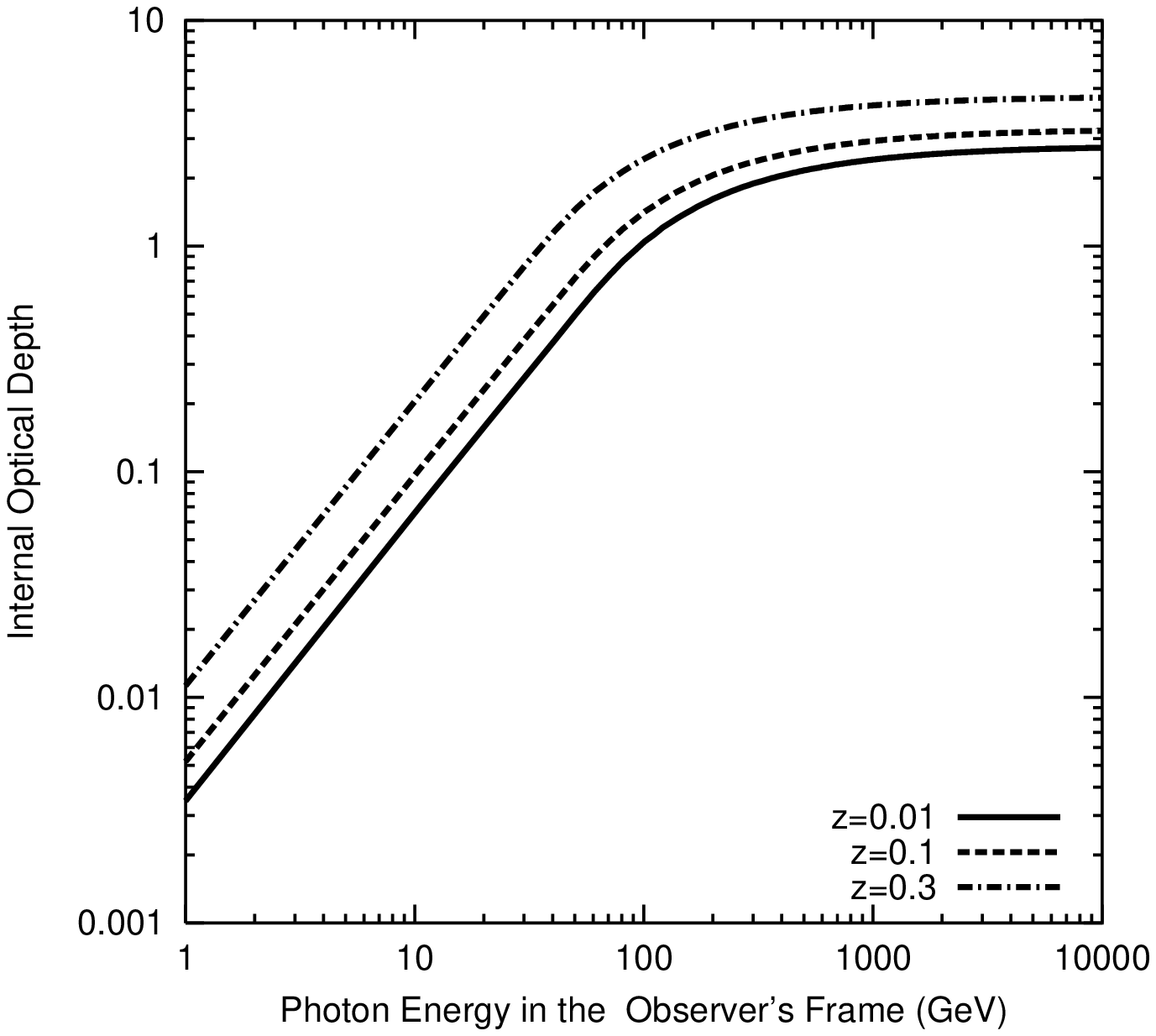,height=9.5cm} 
\caption{Internal optical depth as a function of the photon's energy in 
the observer's frame,  
for various values of the redshift $z$ of the GRB as indicated. Values 
taken for other relevant parameters are: 
$\llowobn=1,\, \beta_l=1,\, \beta_h=2.25,\, 
\epsilon^\ob_b=0.5\mev,\, \tvarob=0.5\sec,\, {\rm and}\,\, \Gamma=300\,.$ 
}
\end{figure}

\newpage 
\begin{figure}
\epsfig{figure=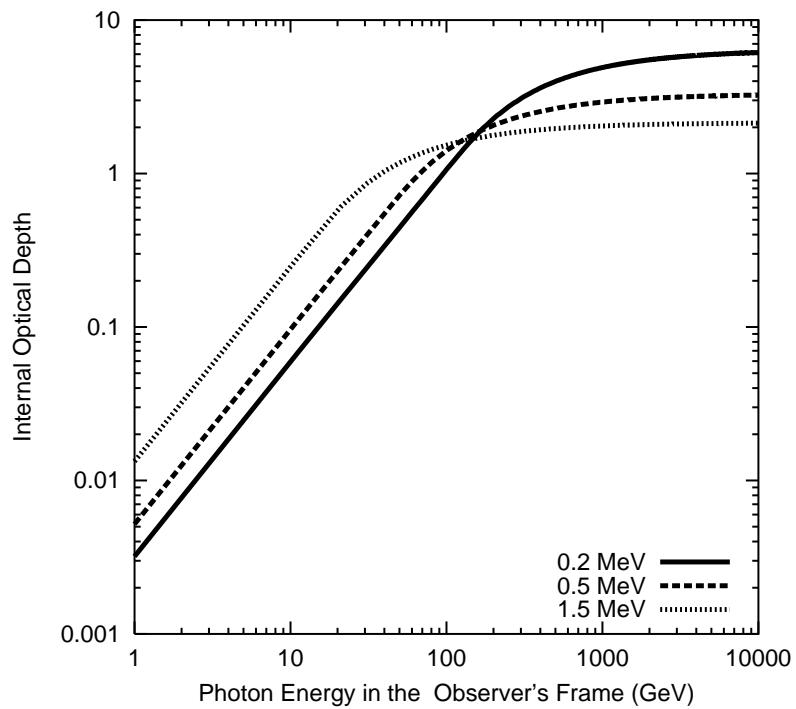,height=9.5cm} 
\caption{Internal optical depth as a function of the photon's energy in 
the observer's frame,  
for various values of the break energy of the observed 
BATSE spectrum, $\epsilon^\ob_b$, as indicated. Values taken for other 
relevant parameters are: $\llowobn=1,\, \beta_l=1,\, \beta_h=2.25,\, 
\tvarob=0.5\sec,\, \Gamma=300,\, {\rm and}\,\, z=0.1\,.$ 
}
\end{figure}

\newpage 
\begin{figure}
\epsfig{figure=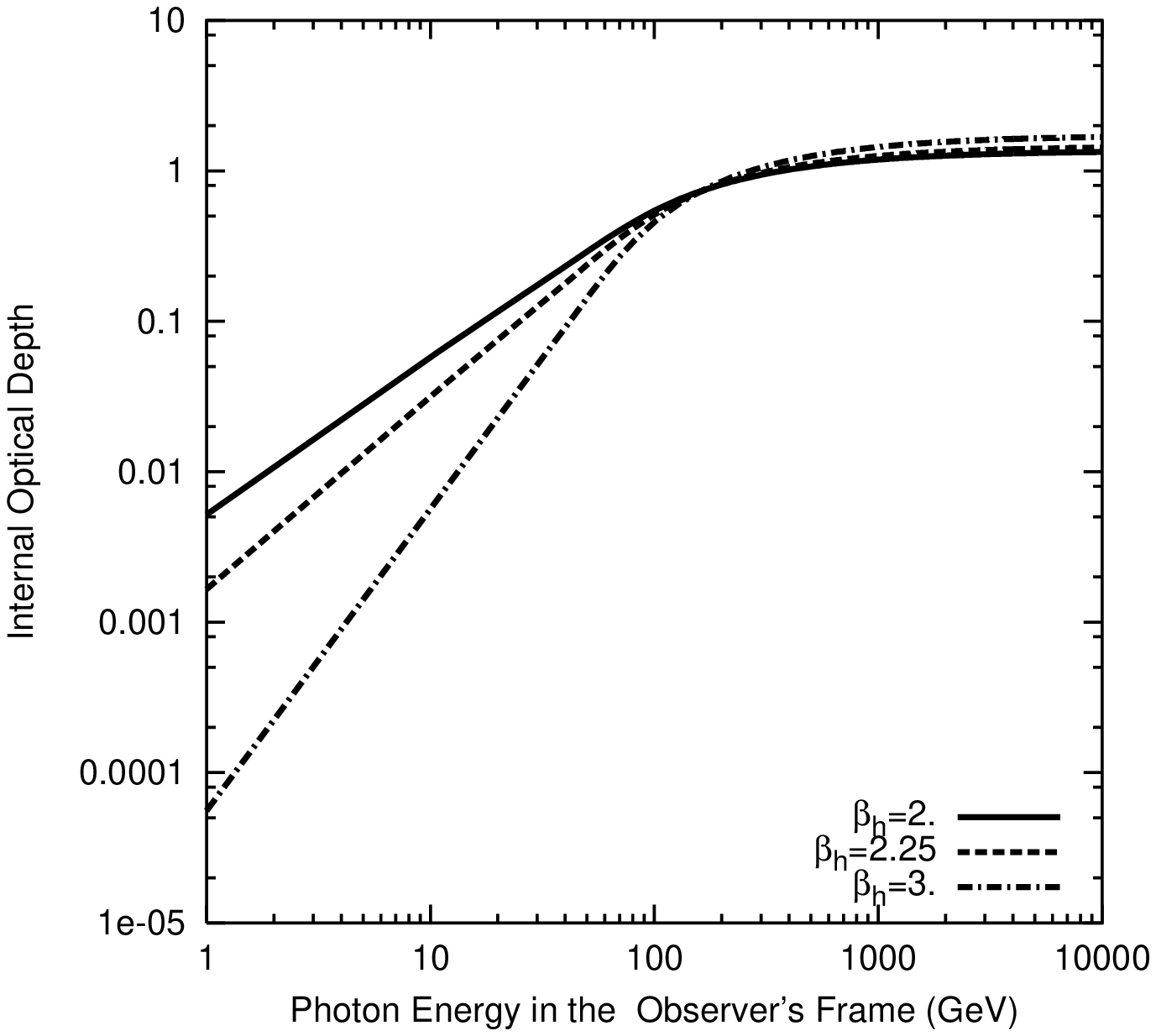,height=9.5cm} 
\caption{Internal optical depth as a function of the photon's energy in 
the observer's frame,  
for various values of $\beta_h$ as indicated. Values taken for other 
relevant parameters are: $\llowobn=1,\, \beta_l=1,\, 
\epsilon^\ob_b=0.5\mev,\, \tvarob=0.5\sec,\, \Gamma=400,\, {\rm and}\,\, 
z=0.3\,.$ 
}
\end{figure}

\newpage 
\begin{figure}
\epsfig{figure=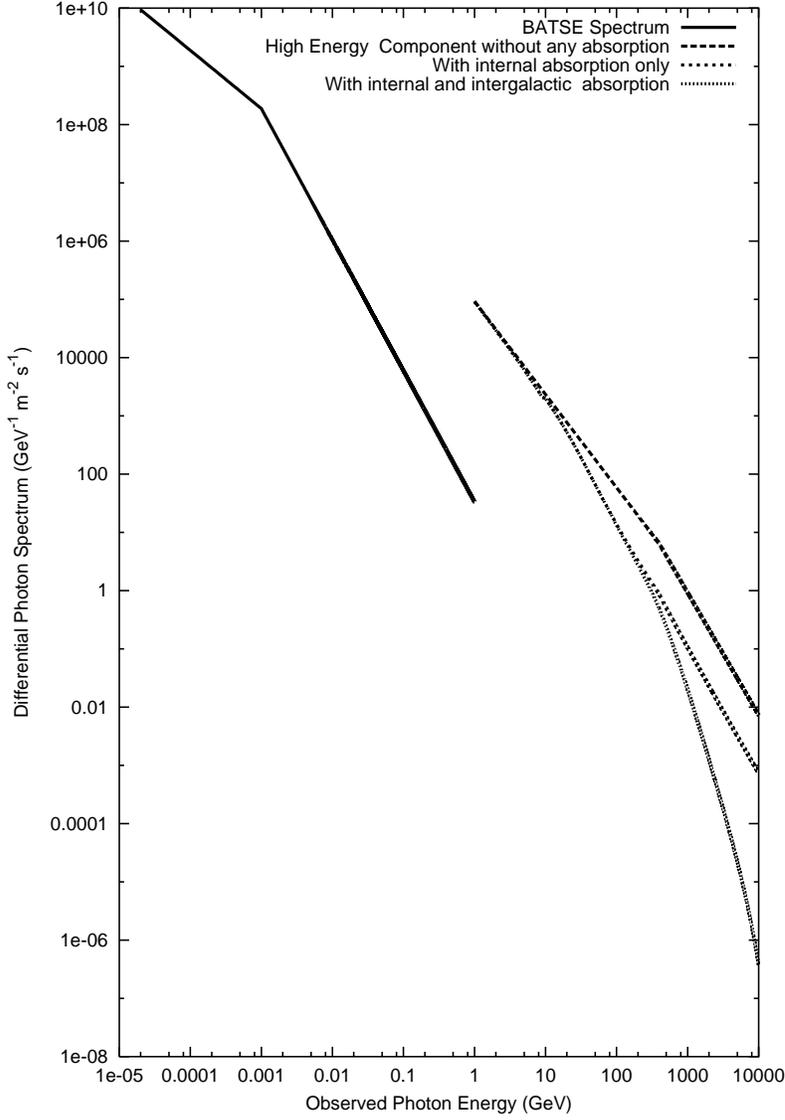,height=15cm} 
\caption{A typical predicted differential spectrum of the high energy 
component of a GRB due to proton-synchrotron process. 
The effects of the internal and intergalactic optical depths 
due to $\gamma\gamma\to e^+e^-$ process are indicated.  
The observed low energy ``BATSE'' spectrum is also shown for 
comparison. Values taken for the various  
relevant parameters are: $\llowobn=1,\, \xi_B=0.5,\, 
\lhighlab=10^{54}\ergs/\sec,\, \alpha_p=2.2,\, 
\beta_l=1,\, \beta_h=2.25,\, \epsilon^\ob_b=1\mev,\, \tvarob=0.5\sec,\, 
\Gamma=300,\, {\rm and}\,\,  
z=0.1\,.$ 
}
\end{figure}

\newpage 
\begin{figure}
\epsfig{figure=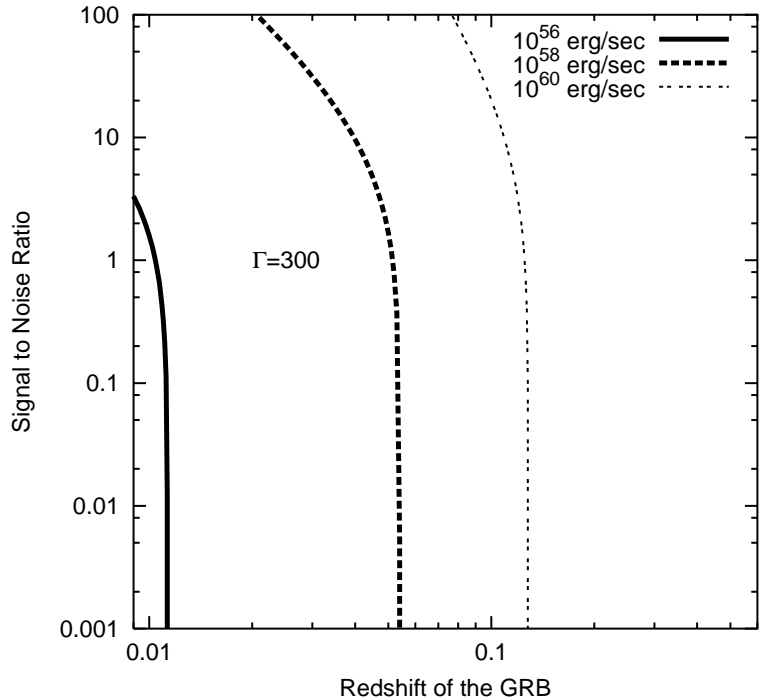,height=9.5cm} 
\caption{Signal to noise ratio in a ICECUBE class detector as a function 
of the redshift of the GRB for various values of the luminosity in 
the high energy component emitted from source, $\lhighlab$, as indicated, 
including the effects of the internal as well as external (intergalactic)  
optical depths. The zenith angle of the GRB source is assumed to be zero, 
and the effective area of the detector for downward-going muons is taken 
to be $\sim 5\times10^5\m^2$. The threshold muon energy (at the surface) 
is taken to be 250 GeV. Values taken for other relevant parameters are: 
$\llowobn=1,\, \xi_B=0.5,\, \alpha_p=2.2,\, 
\beta_l=1,\, \beta_h=2.25,\, \epsilon^\ob_b=1\mev,\, \tvarob=0.5\sec,\, 
\Gamma=300,\, {\rm and}\,\, T^\ob=(1+z)10\sec\,.$ 
}
\end{figure}

\newpage 
\begin{figure}
\epsfig{figure=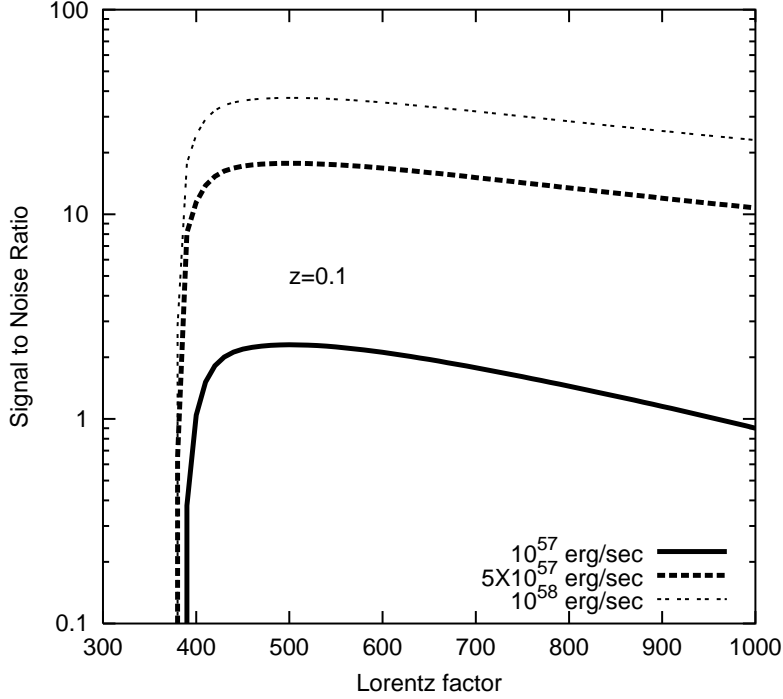,height=9.5cm} 
\caption{Signal to noise ratio in a ICECUBE class detector as a function 
of the Lorentz factor $\Gamma$ of the relativistic outflow 
characterizing the GRB, for various values of the luminosity in 
the high energy component emitted from source, $\lhighlab$, as indicated, 
including the effects of the internal as well as external (intergalactic)  
optical depths. The zenith angle of the GRB source is assumed to be zero, 
and the effective area of the detector for downward-going muons is taken 
to be $\sim 5\times10^5\m^2$. The threshold muon energy (at the surface) 
is taken to be 250 GeV. Values taken for other relevant parameters are: 
$\llowobn=1,\, \xi_B=0.5,\, \alpha_p=2.2,\, 
\beta_l=1,\, \beta_h=2.25,\, \epsilon^\ob_b=1\mev,\, \tvarob=0.5\sec,\, 
z=0.1,\, {\rm and}\,\, T^\ob=(1+z)10\sec\,.$ 
}
\end{figure}

\newpage 
\begin{figure}
\epsfig{figure=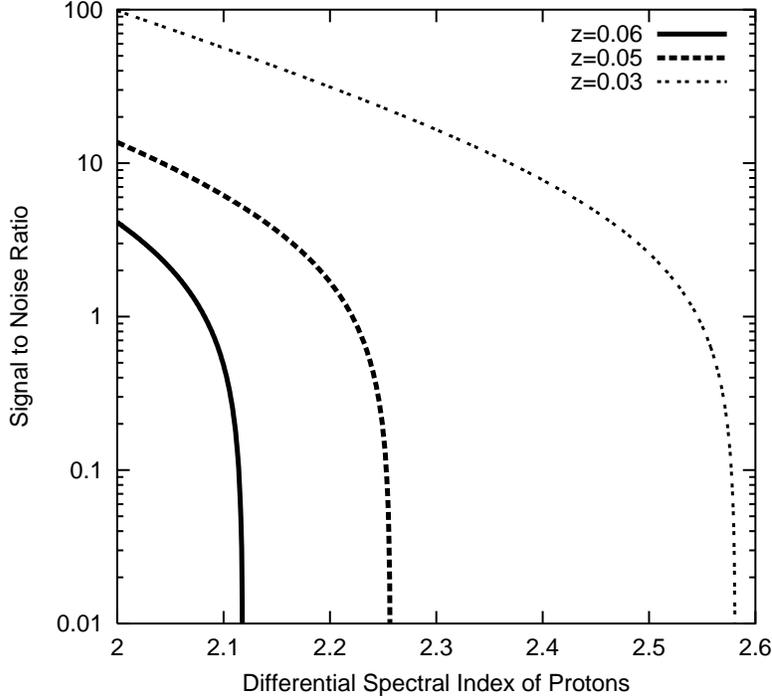,height=9.5cm} 
\caption{Signal to noise ratio in a ICECUBE class detector as a function 
of the power-law index ($\alpha_p$) of the differential spectrum of 
protons accelerated within the GRB source, for various values of the 
redshift $z$ of the GRB as indicated, 
including the effects of the internal as well as external (intergalactic)  
optical depths. The zenith angle of the GRB source is assumed to be zero, 
and the effective area of the detector for downward-going muons is taken 
to be $\sim 5\times10^5\m^2$. The threshold muon energy (at the surface) 
is taken to be 250 GeV. Values taken for other relevant parameters are: 
$\llowobn=1,\, \beta_l=1,\, \beta_h=2.25,\, \epsilon^\ob_b=1\mev,\, 
\xi_B=0.5,\, \tvarob=0.5\sec,\, \Gamma=300,\,  
\lhighlab=10^{58}\ergs/\sec,\, {\rm and}\,\, T^\ob=(1+z)10\sec\,.$ 
}
\end{figure}

\newpage 
\begin{figure}
\epsfig{figure=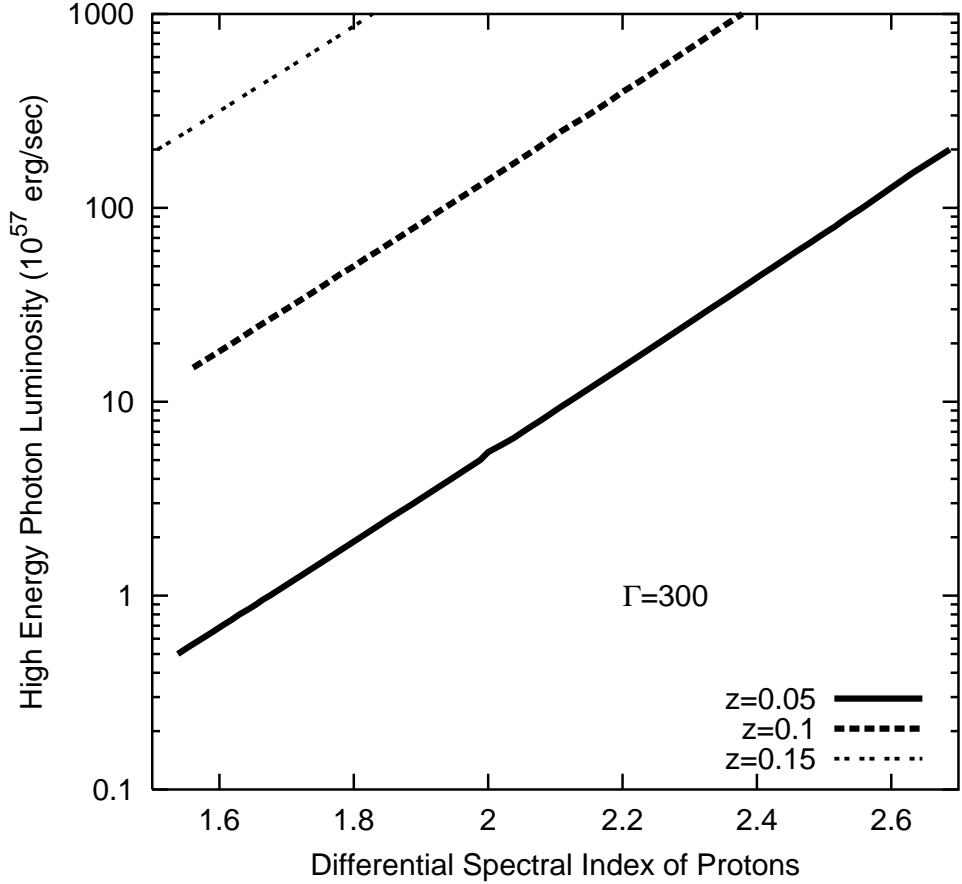,height=12cm} 
\caption{The minimum luminosity in the high energy component, $\lhighlab$, 
required for detection with a signal to noise ratio of 5 or larger in a 
ICECUBE class detector, as a function 
of the power-law index ($\alpha_p$) of the differential spectrum of 
protons accelerated within the GRB source, for various values of the 
redshift $z$ of the GRB as indicated, 
including the effects of the internal as well as external (intergalactic)  
optical depths. The zenith angle of the GRB source is assumed to be zero, 
and the effective area of the detector for downward-going muons is taken 
to be $\sim 5\times10^5\m^2$. The threshold muon energy (at the surface) 
is taken to be 250 GeV. Values taken for other relevant parameters are: 
$\llowobn=1,\, \beta_l=1,\, \beta_h=2.25,\, \epsilon^\ob_b=1\mev,\, 
\xi_B=0.5,\, \tvarob=0.5\sec,\, \Gamma=300,\, {\rm and}\,\, 
T^\ob=(1+z)10\sec\,.$ 
}
\end{figure}
\end{document}